\documentclass[sigconf,nonacm]{aamas}

\usepackage{enumitem}  
\usepackage{balance} 
\usepackage{xspace}

\newcommand{\algoName}{\textsc{MARL-BC}\xspace}

\newlength{\parvspace}
\newlength{\secvspace}
\newlength{\Secvspace}
\setcopyright{none}

\title[AAMAS-2026 Formatting Instructions]{Heterogeneous RBCs via \\ Deep Multi-Agent Reinforcement Learning}

\author{Federico Gabriele}
\affiliation{
\institution{Università La Sapienza di Roma}
\country{Italy}
}
\authornote{Work done during an internship at Banca d'Italia$^{\ddag}$.}
\orcid{0009-0002-1327-8143}

\author{Aldo Glielmo}
\affiliation{%
\institution{Banca d'Italia$^{\ddag}$}
  \country{Italy}
  }
  \authornote{aldo.glielmo@bancaditalia.it, marco.taboga@bancaditalia.it. \\ $^\ddag$The views and opinions expressed in this paper are those of the authors and do not necessarily reflect the official policy or position of Banca d’Italia. \vspace{0.15cm} \\
  This article was published in the proceedings  of the \emph{25th International Conference on Autonomous Agents and Multiagent Systems (AAMAS 2026)}, and is also available at \url{https://doi.org/10.65109/VZHC1838}.
  }
\orcid{0000-0002-4737-2878} 

\author{Marco Taboga}
\affiliation{%
\institution{Banca d'Italia$^{\ddag}$}
  \country{Italy}
  }
\authornotemark[2]
\orcid{0000-0002-5611-5910}

\begin{abstract}
Current macroeconomic models with agent heterogeneity can be broadly divided into two main groups. 
Heterogeneous-agent general equilibrium (GE) models, such as those based on Heterogeneous Agent New Keynesian (HANK) or Krusell--Smith (KS) approaches, rely on GE and `rational expectations', somewhat unrealistic assumptions that make the models very computationally cumbersome, which in turn limits the amount of heterogeneity that can be modelled.
In contrast, agent-based models (ABMs) can flexibly encompass a large number of arbitrarily heterogeneous agents, but typically require the specification of explicit behavioural rules, which can lead to a lengthy trial-and-error model-development process.
To address these limitations, we introduce \algoName, a framework that integrates deep multi-agent reinforcement learning (MARL) with real business cycle (RBC) models.
We demonstrate that \algoName can: (1) recover textbook RBC results when using a single agent; (2) recover the results of the mean-field KS model using a large number of identical agents; and (3) effectively simulate rich heterogeneity among agents, a hard task for traditional GE approaches.
Our framework can be thought of as an ABM if used with a variety of heterogeneous interacting agents, and can reproduce GE results in limit cases.
As such, it is a step towards a synthesis of these often opposed modelling paradigms.
\end{abstract}


\newcommand{\BibTeX}{\rm B\kern-.05em{\sc i\kern-.025em b}\kern-.08em\TeX}

\begin{document}


\pagestyle{fancy}
\fancyhead{}


\maketitle 


\vspace{\Secvspace}
\section{Introduction}

\begin{figure*}[ht]
  \includegraphics[width=0.99 \linewidth]{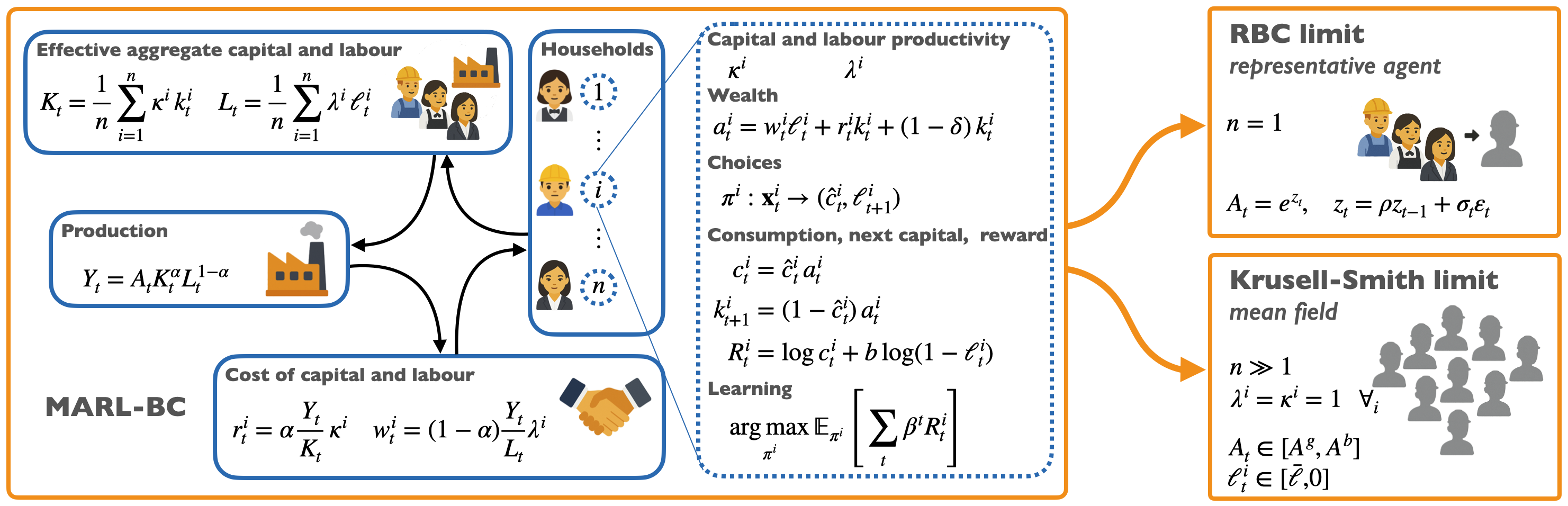}
  \caption{
    The \algoName framework and two limit cases. 
    \normalfont
    The \textbf{left} orange square contains a schematic illustration of the \algoName framework. 
    A population of $n$ heterogeneous RL households, $i=1, \dots, n$, possesses distinct productivities of capital $\kappa_i$ and of labour $\lambda_i$, and provides capital $k^i_t$ and labour $\ell^i_t$.
    These individual inputs aggregate into total capital $K_t$ and total labour $L_t$, which drive production $Y_t$ via a Cobb-Douglas function.
    The individual interest rates $r^i_t$ and the wages $w^i_t$ are assumed to be proportional to the corresponding individual productivities.
    Wealth $a^i_t$ is given by past depreciated capital $(1-\delta)k^i_t$, earned wages $w^i_t \ell^i_t$, and earned interest $r^i_t k^i_t$.
    Each agent decides the fraction $\hat{c}^i_t$ of wealth to consume and the amount of labour to provide.
    These choices determine the instantaneous reward $R^i_t$, which the agent learns to maximise in the long run.
    %
    The two \textbf{right} orange boxes represent classic macroeconomic models as limiting cases of our framework.
    The RBC model (top) is obtained using a representative agent ($n = 1$) and an AR(1) process for the technology shocks $A_t$.
    The Krusell–Smith mean-field model (bottom) is obtained using many ex-ante identical agents and technology and labour switching between discrete values.
    }
    \label{fig:illustration}
\end{figure*}

Macroeconomic modelling has traditionally relied on General Equilibrium (GE) models employing representative agents, such as the real business cycle (RBC) and new Keynesian (NK) models.
However, a well-known limitation of representative-agent models is their inability to account for agent heterogeneity \cite{Kirman1992,stoker1993empirical}.

To capture heterogeneity, two main modelling approaches emerged in macroeconomics: (i) heterogeneous-agent GE models, and (ii) agent-based models (ABMs). 
The first approach embeds heterogeneity directly into a GE framework, typically extending a representative agent model through a `mean-field' approach.
In this setting, individuals act given aggregate quantities, which act as a `mean field', in analogy with physics.
Influential early examples aimed at introducing heterogeneous households into the RBC setting are those of \citet{Aiyagari1994} and of \citet{KrusellSmith1998}.
More recently, Heterogeneous Agent New Keynesian (HANK) models have incorporated similar extensions into NK models \cite{KaplanMollViolante2018,auclert2025fiscal}.
These GE heterogeneous-agent models come at a high cost, as each agent is assumed to have `rational expectations' and needs to keep track of the entire wealth or income distribution as state variable to make any decision, an assumption that can be thought of as both unrealistic and too computationally burdensome \cite{moll2024trouble}.
Furthermore, the high computational costs involved significantly restrict the degree of heterogeneity achievable with GE models.
For example, often heterogeneity appears `ex-post', meaning that all agents are initially identical and differentiate only as a result of the individual shocks they undergo during the simulation.
The second major approach is agent-based modelling.
ABMs abandon the representative-agent and rational expectations assumptions entirely.
They simulate economies `from the bottom up' with many interacting agents that are both ex-ante and ex-post heterogeneous, have no rational expectations, and instead are endowed with boundedly rational heuristic decision rules \cite{FarmerFoley2009, dawid2018agent,farmer2024making,axtell2025agent,borsos2025agent}.
This flexibility of ABMs is also the reason why they face the strongest criticism.
Because in ABMs the modeller typically needs to decide the agents' behaviour directly, it can be difficult to correctly navigate the arbitrariness in their specification and pin down realistic rules~\cite{windrum2007empirical,benedetti2022black}.

Only very recently, researchers have begun exploring reinforcement learning (RL) \cite{sutton1998reinforcement} – and especially multi-agent reinforcement learning (MARL) \cite{albrecht2024multi} – as a novel way to model heterogeneous agents in macroeconomics \cite{atashbar2022deep,cook2025reinforcement}. 
RL agents learn optimal (or nearly optimal) behaviours through repeated interaction with an environment to maximise their reward signals over time.
The RL learning paradigm seems to offer a natural synthesis between the extremes of GE models and ABMs: agents can be boundedly rational and diverse, yet their behaviours emerge endogenously from a principled optimisation process (learning to maximise rewards), so that the modeller can avoid searching directly for a behavioural rule \cite{brusatin2024simulating}.
Moreover, modern deep RL techniques can solve complex, high-dimensional and nonlinear optimisation problems, which would be difficult to tackle with traditional methods.

In this work, we follow this direction and show how deep RL techniques can bridge the gap between ABMs and heterogeneous-agent GE models.
Our contributions are as follows:
\begin{itemize}[leftmargin=0.7cm]
    \item We develop the Multi-Agent Reinforcement Learning Business Cycle (\algoName), a MARL-based framework that extends the classic RBC model with multiple households characterised by rich and flexible heterogeneity.
    \item We demonstrate that the training of \algoName is computationally feasible, employing state-of-the-art RL algorithms such as Proximal Policy Optimisation (PPO) or Soft Actor Critic (SAC) or Deep Deterministic Policy Gradient (DDPG).
    \item We show that \algoName can recover classic textbook RBC results when using a single agent.
    \item We further show that \algoName can recover the mean-field Krusell--Smith model when using a large population of ex-ante identical agents.
    \item Finally, we illustrate the capability of the framework to simulate richer heterogeneity among agents, impossible to achieve with a GE approach.
\end{itemize}
Our results demonstrate the value and relevance of the \algoName framework for a growing community of researchers in both economics and computer science, thus paving the way for numerous applications and future investigations.

\vspace{\parvspace}
\noindent
\textbf{Related work.}
The idea of using RL as a model for realistic human decision making is attracting increasing research and development interest for economic and financial simulations \cite{tilbury2022reinforcement,atashbar2022deep}.
Influential early contributions that sparked interest in this direction were the studies in \cite{johanson2022emergent} and \cite{zheng2022ai}, where deep multi-agent RL was employed to simulate emergent economic behaviours within simplified toy economies.
Initially, its adoption occurred mostly within the financial sector, where RL was successfully applied to model trading strategies in different markets \cite{ardon2021towards,mascioli2024financial,gao2023deeper,Evans2024,bacaloni2025natural}, and has led to the development of specialised open-source platforms \cite{amrouni2021abides,phantom2023}.
In macroeconomics, recent literature has begun exploring RL techniques to extend classic GE frameworks \cite{cook2025reinforcement}. 
For instance, \cite{chen2021deep,hinterlang2021optimal} use RL for monetary policy, while \cite{CurryEtAl2023,curry2022analyzing,MiEtAl2023,dwarakanath2024tax} employ MARL for tax policy optimisation. 
In \cite{AtashbarShi2023} and \cite{yang2025structural}, RL is used to solve representative-agent and heterogeneous GE models respectively, while in \cite{hill2021solving} it is leveraged to solve a hybrid model that combines features of both macro ABMs and GE models.
Parallel studies focused on extending RL with ABMs. 
For instance, in \cite{brusatin2024simulating}, the authors incorporate RL firms within a standard ABM, while in \cite{glielmo2023reinforcement} the authors use RL to enhance the calibration of ABMs.
In \cite{agrawal2025robust}, RL is used to design robust macroeconomic policies resilient to model misspecification and uncertainty.
Similarly, \cite{Evans2025} proposes a general framework combining adaptive agent behaviours through RL within ABMs.
Our work is also related to another line of ongoing research which aims to tackle the curse of dimensionality of GE models with deep learning, but without using RL \citep{fernandez2024taming,fernandez2025deep}.
Finally, our work can contribute to research studying the effects of bounded rationality, typically done without RL, in existing ABMs or GE models \cite{dosi2020rational,asano2021emergent}.

\vspace{\parvspace}
\noindent
\textbf{Significance.}
Our work stands out clearly in comparison with the described literature for the following reasons.
Existing literature at the \emph{economics} end of the spectrum has mostly focused on single-agent RL, showing that it can solve representative‑agent GE models.
However, research in economics has \emph{not} explored how to build more expressive models using a multi-agent RL approach.
Conversely, the \emph{computer‑science} literature has experimented with multi-agent RL, showing that it can give rise to rich emergent economic behaviours.
However, research in computer science has mostly overlooked fundamental models from macroeconomics, and it has \emph{not} convincingly shown how multi-agent RL can extend such models.
In our view, this has created a clear knowledge gap between the two disciplines, which currently impedes communication and mutually beneficial exchanges of ideas.
For example, this implied that, on the one hand, economic modelling could not leverage the full potential of multi-agent RL as a viable strategy for extending its traditional modelling frameworks and, on the other hand, that research and development in computer science could not progress rapidly towards the construction of frameworks of actual relevance for the economics discipline.

Our work represents a clear step forward in bridging this knowledge gap, as it provides a foundation for linking research in two disciplines.
For this reason, we believe it has significant potential for stimulating fruitful mutual exchanges.
Specifically, we bridge the two research lines by proposing a framework that (i) exactly reproduces the representative‑agent RBC solution when $n=1$, (ii) converges to the KS mean‑field equilibrium as $n \gg 1$, and (iii) extends beyond both limits to easily accommodate a high degree of discrete heterogeneity that GE methods cannot easily handle.

The rest of this work is structured as follows. 
In Sec. \ref{sec:model} we describe the \algoName framework.
In Sec. \ref{sec:experimental_setup}, we provide details on the training and testing procedure we followed for the numerical experiments.
In Sec. \ref{sec:results}, we describe the results of our numerical experiments, dividing the discussion into three parts: the replication of textbook RBC results, the replication of mean-field Krusell--Smith results, and the illustration of the full capabilities of our framework.
Finally, in Sec. \ref{sec:conclusions}, we conclude.

\vspace{\Secvspace}
\section{The \algoName framework}
\label{sec:model}

We describe here the Multi-Agent Reinforcement Learning Business Cycle (\algoName) framework, in its general form. 
This will allow us to analyse the traditional models, such as the textbook RBCs or the Krusell--Smith, simply as limiting cases. 
Our framework and the mentioned limiting cases are illustrated in Figure~\ref{fig:illustration}.
Specifically, the orange box in the figure contains the general \algoName framework. The framework is composed of two pieces: the heterogeneous RBC environment, depicted as the four solid blue boxes, is described in Sec.~\ref{subsec:model_environment}, while the RL household problem, depicted as the dashed blue box, is described in Sec.~\ref{subsec:model_rl_household}.

\vspace{\secvspace}
\subsection{The heterogeneous RBC environment}
\label{subsec:model_environment}
The model consists of $n$ types of households $i=1,\dots,n$ and a single firm. 
At each time $t$, each type of household (which we will also refer to as a `household') can own a different amount of capital $k^i_t$ and can provide a different amount of labour $\ell^i_t$.
Furthermore, each agent has a different productivity of capital $\kappa^{i}$ and of labour $\lambda^{i}$.
We assume here that capital and labour productivities are fixed in time, but the model can be easily extended to allow them to vary.

\vspace{\parvspace}
\noindent
\textbf{Effective aggregate capital and labour.}
Households' capital and labour are assumed to give rise to the effective aggregate capital $K_t$ and labour $L_t$ as
\begin{equation}\label{eq:agg}
\begin{aligned}
 K_t &= \frac{1}{n}\sum_{i=1}^{n} \kappa^i\, k^i_t,
 &\qquad
 L_t &= \frac{1}{n}\sum_{i=1}^{n} \lambda^i \, \ell^i_t,
\end{aligned}
\end{equation}
which are averages of the households' variables, weighted by the corresponding productivities.
Although these are formally written averages, we take the population mass to be one (a convention that also ensures consistency with standard GE models), so averages coincide with aggregates, quantities are expressed on a per capita basis, and the firm's conditions in Eq. \eqref{eq:returns_and_wages} apply.

\vspace{\parvspace}
\noindent
\textbf{Production.}
Given aggregate capital and labour, production $Y_t$ is given by a Cobb-Douglas function
\begin{equation}
Y_t = A_t K_t^\alpha L_t^{1-\alpha}.
\label{eq:production}
\end{equation}
with output elasticity of capital $\alpha$, and technology $A_t$.

\vspace{\parvspace}
\noindent
\textbf{Cost of capital and labour.}
We assume that labour and capital markets are perfectly competitive. Therefore, wages $w^i_t$ and interest rates $r^i_t$ are proportional to the marginal productivities of capital and labour respectively,
\begin{equation} r^i_t = \alpha\frac{Y_t}{K_t}\, \kappa_i, \quad \quad \quad w^i_t = (1-\alpha)\frac{Y_t}{L_t} \lambda_i. \label{eq:returns_and_wages} \end{equation}

\vspace{\parvspace}
\noindent
\textbf{Households' wealth.}
Each household receives wages and interest as compensation for its labour and capital contributions to aggregate output while it faces depreciation of its capital stock at a fixed rate $\delta$. Hence, the wealth $a^i_{t}$ of household $i$ at time $t$ is
\begin{equation}\label{eq:kap_motion}
    a^i_{t} = w^i_t\ell^i_t + r^i_t k^i_t + (1-\delta)\,k^i_t,
\end{equation}
which is the sum of labour earnings $w^i_t\ell^i_t$, returns on capital $r^i_tk^i_t$, and the depreciated  stock of capital $(1-\delta)k^i_t$.

\vspace{\secvspace}
\subsection{The RL households}
\label{subsec:model_rl_household}
\vspace{\parvspace}
\noindent
\textbf{Action space.}
The RL households' action at each time step $t$ is a tuple $(\hat{c}^i_t, \ell^i_{t+1})$ of two continuous numbers, the consumption fraction $\hat{c}^i_t$ and the labour $\ell^i_{t+1}$.
While theoretically both of these fall in the range $(0, 1)$, for numerical stability we clip them in the range $(0.01, 0.99)$.

The consumption fraction $\hat{c}^i_t$ determines $c^i_t$, which is the amount of the household's wealth spent on consumption
\begin{equation}
    c^i_t = \hat{c}^i_t \, a^i_t ,
\end{equation}
as well as the amount dedicated to investments in next-step capital 
\begin{equation}
    k^i_{t+1} = (1-\hat{c}^i_t) \, a^i_t .
\end{equation}
By choosing $\hat{c}^i_t$ and $\ell^i_{t+1}$, the households need to balance current consumption and leisure with future earnings, as will be made clearer in the rest of this section.

\begin{figure}[t]
  \includegraphics[width=0.90 \linewidth]{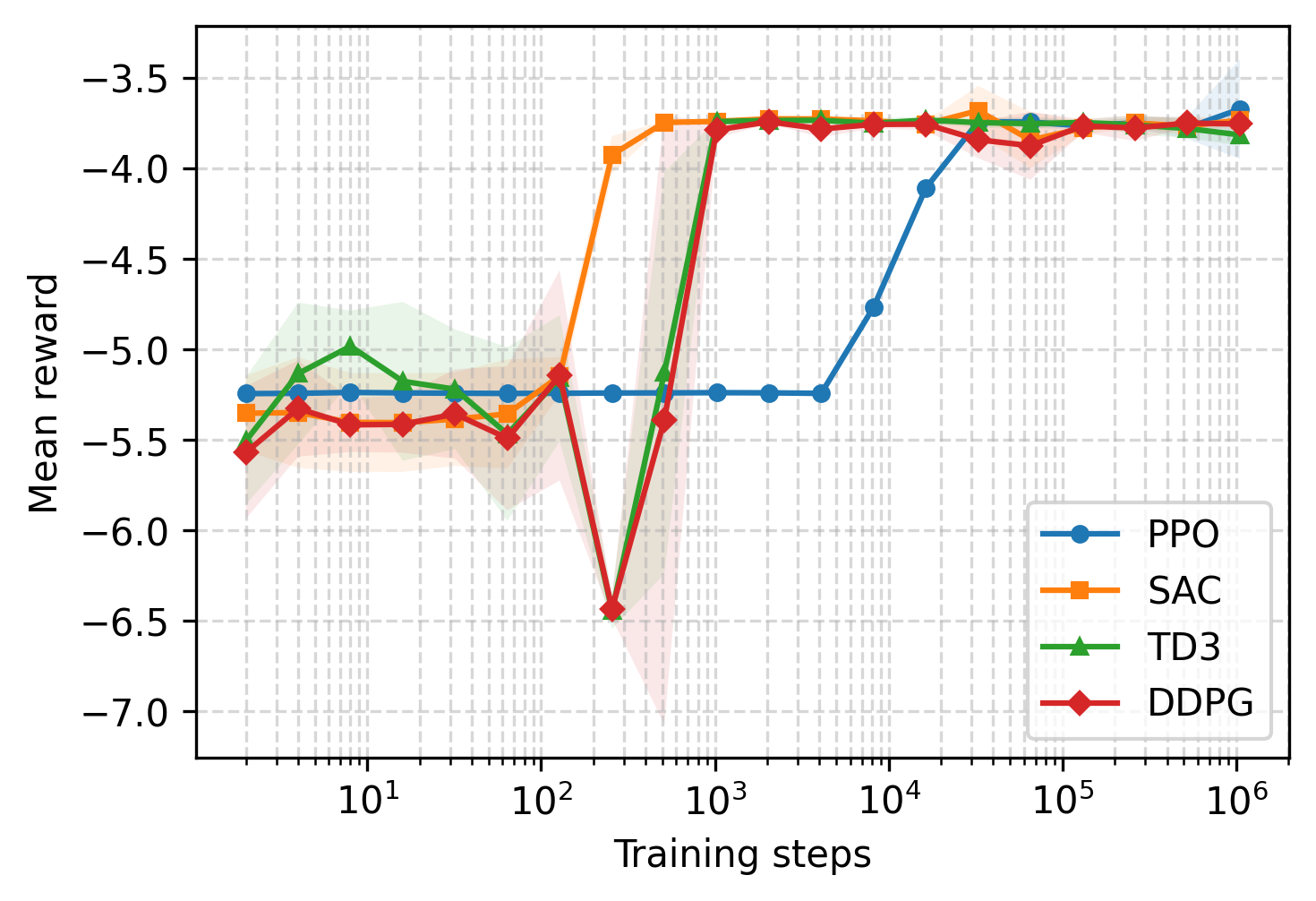}
  \caption{
  State-of-the-art RL schemes can solve the household optimisation problem.
  \normalfont
  Mean rewards obtained by different RL schemes as a function of the number of training steps.
  All schemes considered succeed in learning a policy that maximises cumulative rewards. SAC, DDPG and TD3 are seen to perform significantly better than PPO in terms of sample efficiency. However, PPO's greater computational efficiency (not shown in this graph) can make also this method a good candidate for the task.
  }
  \label{fig:learning_curves}
\end{figure}

\vspace{\parvspace}
\noindent
\textbf{Observation space.}
The observation space $\mathbf{x}^i_t$ of the RL household $i$ can flexibly contain any information on the state of the agent and on the state of the economy that is deemed important for the agent's decisions.
A rather general choice, coherent with standard economic models, would be
\begin{equation}
    \mathbf{x}^i_t =(k^i_t, K_t,  \ell^i_{t}, L_{t}, A_t, \kappa^i, \lambda^i),
\end{equation}
containing the values of individual and aggregate capital, past individual and aggregate labour, and technology.
In practice, not all of these variables are needed to obtain interesting dynamics or to reproduce traditional macroeconomic models, and, in fact, we will sometimes select a subset as specified in the results section.

\vspace{\parvspace}
\noindent
\textbf{Reward.}
Each RL household learns a deterministic policy 
\begin{equation}
    \pi^i: \mathbf{x}^i_t \rightarrow (\hat{c}^i_t, \ell^i_{t+1})
\end{equation}
giving consumption fraction $\hat{c}^i_t$ and labour $\ell^i_{t+1}$ as a function of the state of the household and of the economy $\mathbf{x}^i_t$.
The policy $\pi^i$ is learned by maximising $\mathcal{R}^i$, the expected sum of discounted rewards
\begin{equation}
\mathcal{R}^i = \mathbb{E}_{\pi^i} \left[ \sum_t^T \beta^t R^i_t \right],
\end{equation}
where $\beta \in(0,1)$ is the agent’s discount factor and the superscript in $\beta^t$ here indicates an exponentiation to the power $t$, differently from anywhere else in this work, where it is just used as an index.
The reward $R^i_t$ of household $i$ at time $t$ is defined as
\begin{equation}
R^i_t = \log c^i_t + b \log (1- \ell^i_t).
\label{eq:reward}
\end{equation}
It increases with increasing consumption $c^i_t$ and leisure $(1- \ell^i_t)$, with $b>0$ controlling the trade-off between the two. 
Rewards are known as `utilities' in the economic literature.
Furthermore, the specific function in Eq.~\eqref{eq:reward} is very commonly used since the logarithmic utility is, in fact, a special case of the well-known `Constant Relative Risk Aversion' (CRRA) utility function when the coefficient of relative risk aversion equals one \cite{romer2018advanced}.

\vspace{\parvspace}
\noindent
\textbf{Game-theoretic considerations.}
From a game-theoretic perspective, the households can be thought of as participating in a stochastic game with $n$ players $i=1,\dots,n$, a common state, individual actions $(\hat c_t^i,\ell_{t+1}^i)$, a transition kernel induced by Eqs. (\ref{eq:agg}–\ref{eq:kap_motion}), and per-period payoffs $R^i_t$.
This, in turn, ensures the existence of optimal policies for the households assuming a solution concept such as Markov perfect equilibrium \cite{maskin2001markov}. 
However, the corresponding problem is well known to be computationally hard \cite{daskalakis2009complexity}. 
Consequently, although the approach would be theoretically possible, it is impractical due to the high complexity and dimensionality of the underlying model and its strategy space.
In addition, relying on the full information set of the model is an unrealistic assumption, as real-world agents typically lack full observability of their environment. 
In contrast, we train independent learners each having access only to the partial information set $\mathbf x_t^i$ and optimising for a policy $\pi^i$ that approximates a best response to the evolving behaviour of the others. 
In our numerical experiments, we show that for sufficiently large $n$, the resulting MARL scheme reproduces the equilibrium behaviour of~\cite{KrusellSmith1998}. 
However, from a theoretical standpoint, establishing convergence of independent MARL dynamics to mean field equilibria in stochastic games remains a challenging and open problem \cite{yongacoglu2024mean,yardim2023policy}.

\begin{table}[b]
\centering
\begin{tabular}{lll}
\toprule
\textbf{Symbol} & \textbf{Description} & \textbf{Value (RBC} $\mid$ \textbf{KS} $\mid$ \textbf{General)}\\
\midrule
$n$                & Number of households                         & $ 1 \mid 20 \mid 20 $ \\
$T$                & Episode length                           & 500 \\
\addlinespace[4pt]
$\kappa_{i}$       & Capital productivity                 & $ 1 \mid 1 \mid \{0, 0.8, 1, 1.2, 0.98, 1.02\} $ \\
$\lambda_{i}$      & Labour productivity                  & $ 1 \mid 1 \mid \{0.98, 1, 1.02\} $ \\
\addlinespace[4pt]
$A_{t}$            & Technology                     & $ \text{AR(1)} \mid \text{AR(1)} \mid \{\text{KS}, \text{AR(1)}\} $ \\
$\alpha$           & Output elasticity                       & 0.36 \\
$\delta$           & Capital depreciation                    & $ \{ 1, 0.025 \} \mid 0.025 \mid 0.025 $ \\
\addlinespace[4pt]
$\beta$            & Discount factor                      & 0.95 \\
$b $                & Leisure weight                       & $5 \mid 0 \mid \{ 0, 5 \}$ \\
\bottomrule
\end{tabular}%
\caption{Parameters used in the experiments.
\normalfont
When three values are reported, separated by a vertical bar, they refer respectively to the three result categories (RBC, KS, and General) as specified in the main text.
For the AR(1) technology process we use $\rho=0.9$ and $\sigma=0.01$. The KS technology process is described in the main text.
}
\label{tab:params}
\end{table}

\vspace{\Secvspace}
\section{Experimental setup}
\label{sec:experimental_setup}
In our numerical experiments, we first showcase the capabilities of our framework to reproduce the traditional representative agent RBC model and the mean-field Krusell--Smith (KS) model (Sec. \ref{subsec:results_rbc} and \ref{subsec:results_ks}), and we then show the capabilities of the framework in the general form (Sec. \ref{subsec:results_general}).
In the rest of the paper, we refer to these three experimental settings as `RBC', `KS' and `General'. 

\vspace{0.2cm}
\noindent
\textbf{Parameters.}
Table~\ref{tab:params} reports the main parameters used in our numerical experiments, along with their names and symbols.
The values reported are grouped depending on the type of experiment performed, either `RBC', `KS', or `General'. 
If a single value is reported, then the same value has been used for all experiments.
We note here that most of the parameters used are standard for macroeconomic models of these types \cite{romer2018advanced}.

\begin{figure*}[t]
  \centering
  \includegraphics[width=0.33\textwidth]{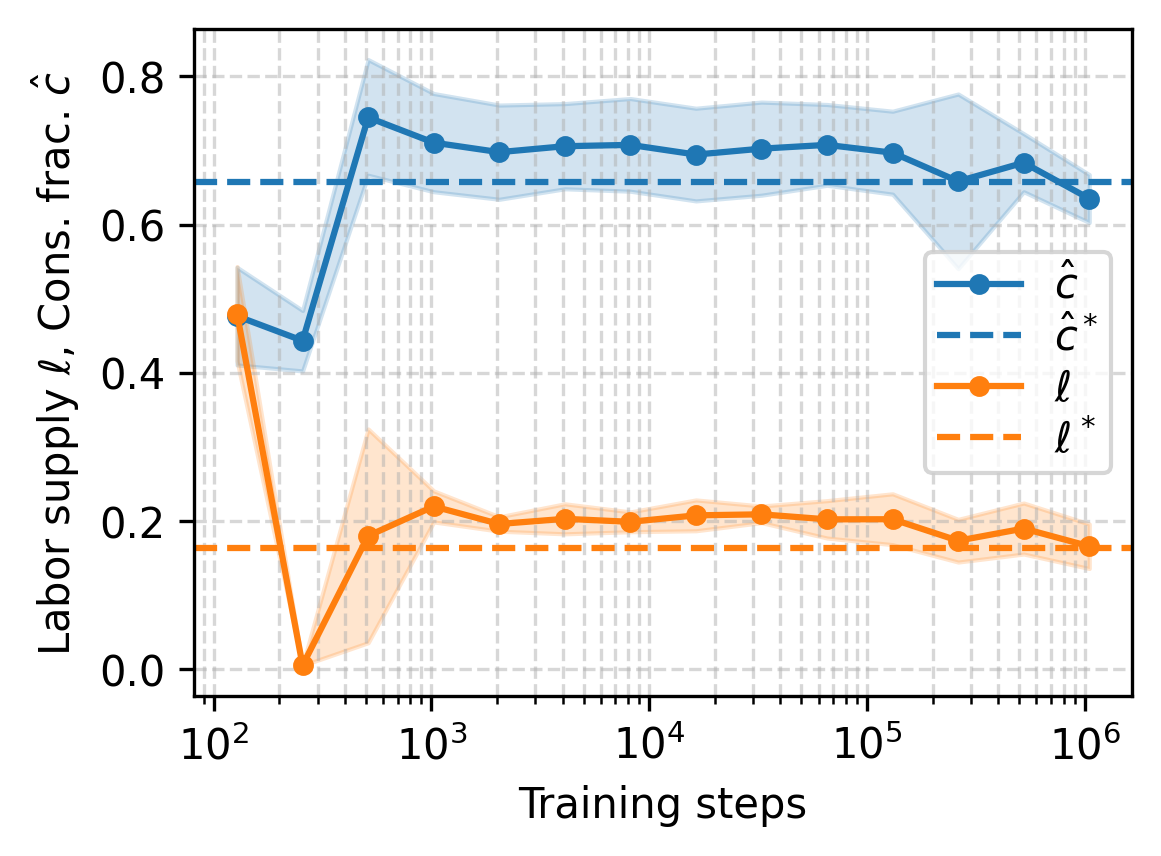}\hfill
  \includegraphics[width=0.33\textwidth]{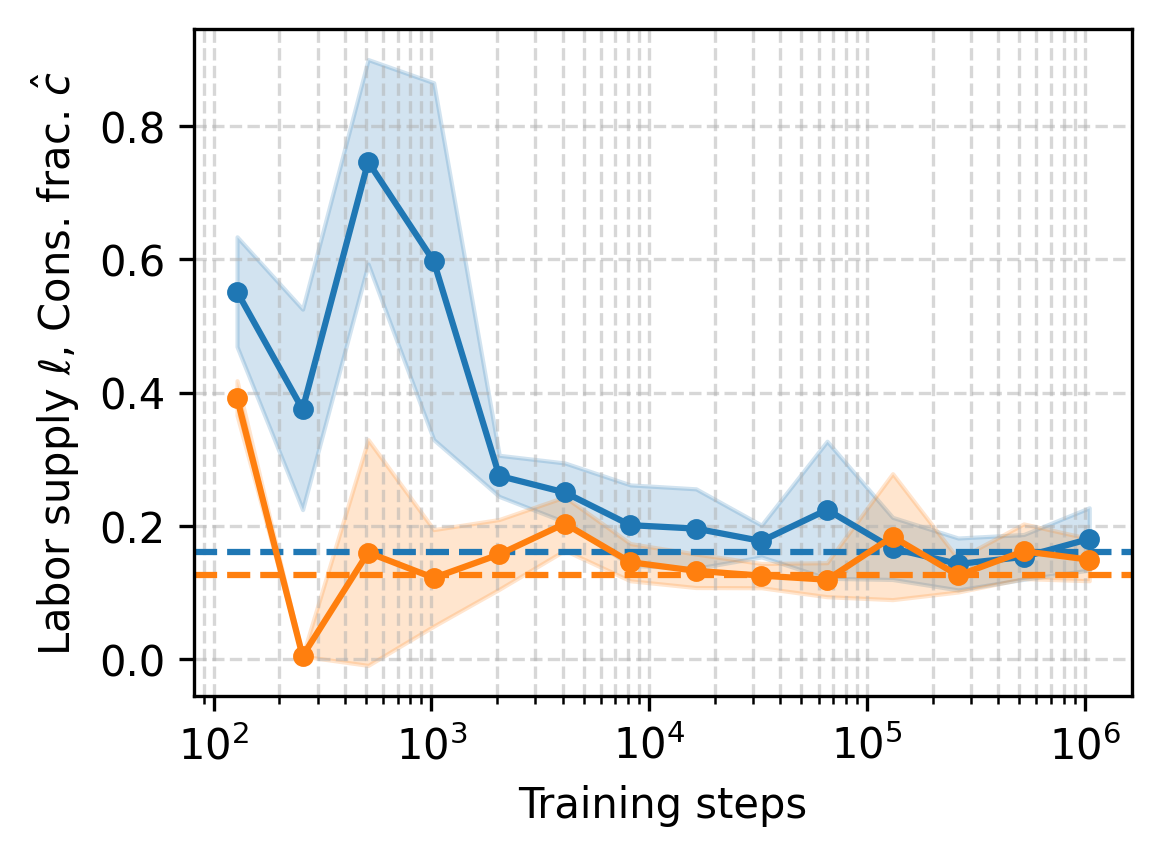}\hfill
  \includegraphics[width=0.33\textwidth]{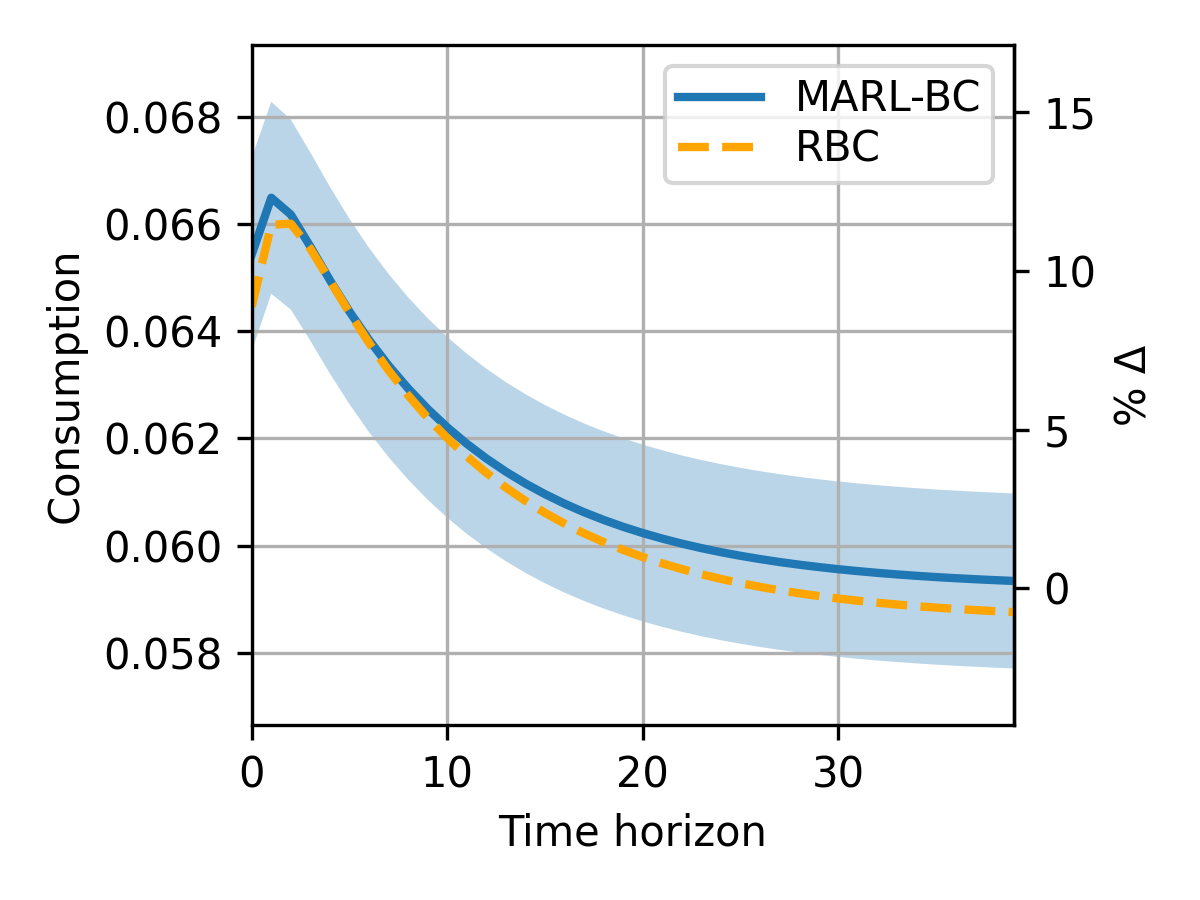}
  \caption{\algoName{} can reproduce representative‐agent results with a single agent.  
  \normalfont
  The \textbf{left} panel shows how the RL household recovers the optimal consumption fraction $\hat{c}$ and labour supply $\ell$ of Eq.~\eqref{eq:textbook_rbc_actions}, for a textbook, analytically solvable, RBC model with full capital depreciation ($\delta=1)$.
  The \textbf{centre} panel shows a similar convergence, but for a more complex model with a partial depreciation ($\delta=0.025$). 
  This model cannot be analytically solved, and the optimal choices are obtained here using numerical tools common in economics.
  Finally, the \textbf{right} panel shows that \algoName is also capable of reproducing the dynamical behaviour of the standard RBC models as captured by an impulse response function.
  The plot shows the impulse response for consumption following a productivity shock of one standard deviation, with the dashed orange line indicating the standard RBC result and the full blue line indicating the equivalent \algoName behaviour.
  }
  \label{fig:rep_agent_results}
\end{figure*}

\vspace{\parvspace}
\noindent
\textbf{Training and testing.}
We implemented \algoName using well-known open-source libraries in Python. 
Specifically, we use the MARL interface of PettingZoo \cite{terry2021pettingzoo}, and use SuperSuit \cite{terry2020supersuit} to leverage RL algorithms from Stable-Baselines3 \cite{raffin2021stable} in an independent MARL fashion.
We consider the following RL algorithms: Deep Deterministic Policy Gradient (DDPG) \cite{LillicrapHPHETS15}, Twin Delayed Deep Deterministic Policy Gradient (TD3) \cite{fujimoto2018addressing}, 
Soft Actor Critic (SAC) \cite{haarnoja2018soft}, Proximal Policy Optimization (PPO) \cite{schulman2017proximal}.
We use default hyperparameters as detailed in \cite[Appendix]{gabriele2025heterogeneous_ARXIV}. 

To facilitate quick convergence and scalability, we adopt the standard MARL paradigm of parameter sharing \cite{albrecht2024multi}.
In practice, this means a single neural network represents all the agents.
Importantly, this paradigm does \emph{not} force an identical behaviour for all agents, since the different agents' characteristics (such as the individual productivities of labour and capital) appear as observations for the neural network, which can thus encode completely different behaviours.
Parameter sharing significantly reduces the computational burden, improves sample efficiency and scalability, and can help to stabilise training.
Furthermore, it can also be thought of as a stylised form of `social learning' \cite{whalen2018sensitivity,ndousse2021emergent} since information discovered by an agent is immediately made available to similar agents via shared parameters.

We train all algorithms for $10^6$ steps in the single-agent environment.
For the multi-agent runs, we perform $10^5$ per-agent updates, corresponding to $n \cdot 10^5$ total steps, sometimes more if needed for convergence (as detailed in \cite[Appendix]{gabriele2025heterogeneous_ARXIV}).
Unless otherwise stated, we report mean and standard deviation over five different training runs.
The code for the \algoName environment used to perform our experiments is available in open source at \url{https://github.com/fedegabriele/MARL-BC}.

\vspace{\Secvspace}
\section{Results}
\label{sec:results}

\begin{figure*}[t]
  \centering
  \includegraphics[width=0.33\textwidth, trim={0.19cm 0.1cm 0.19cm 0.2cm}, clip]{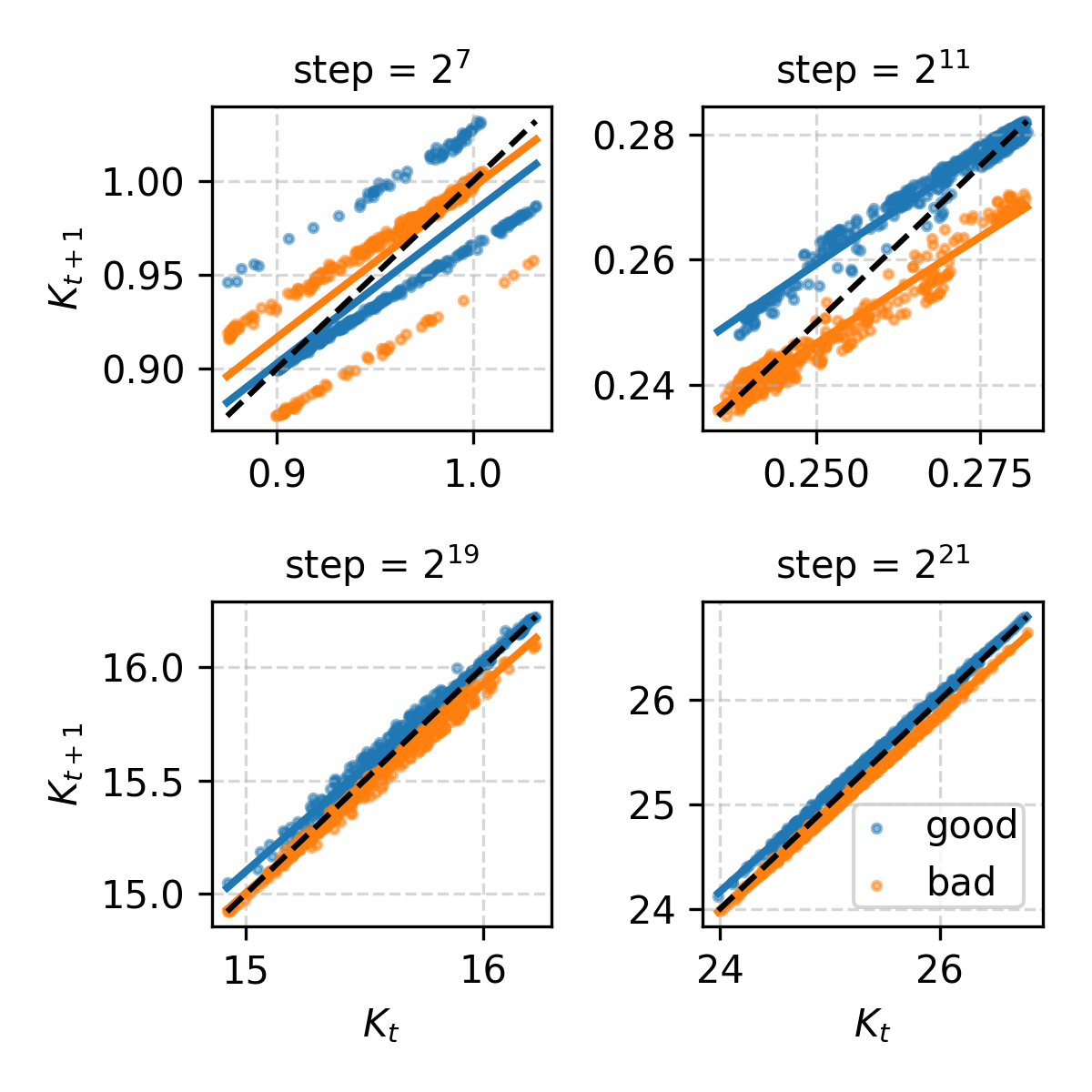}\hfill
  \includegraphics[width=0.33\textwidth]{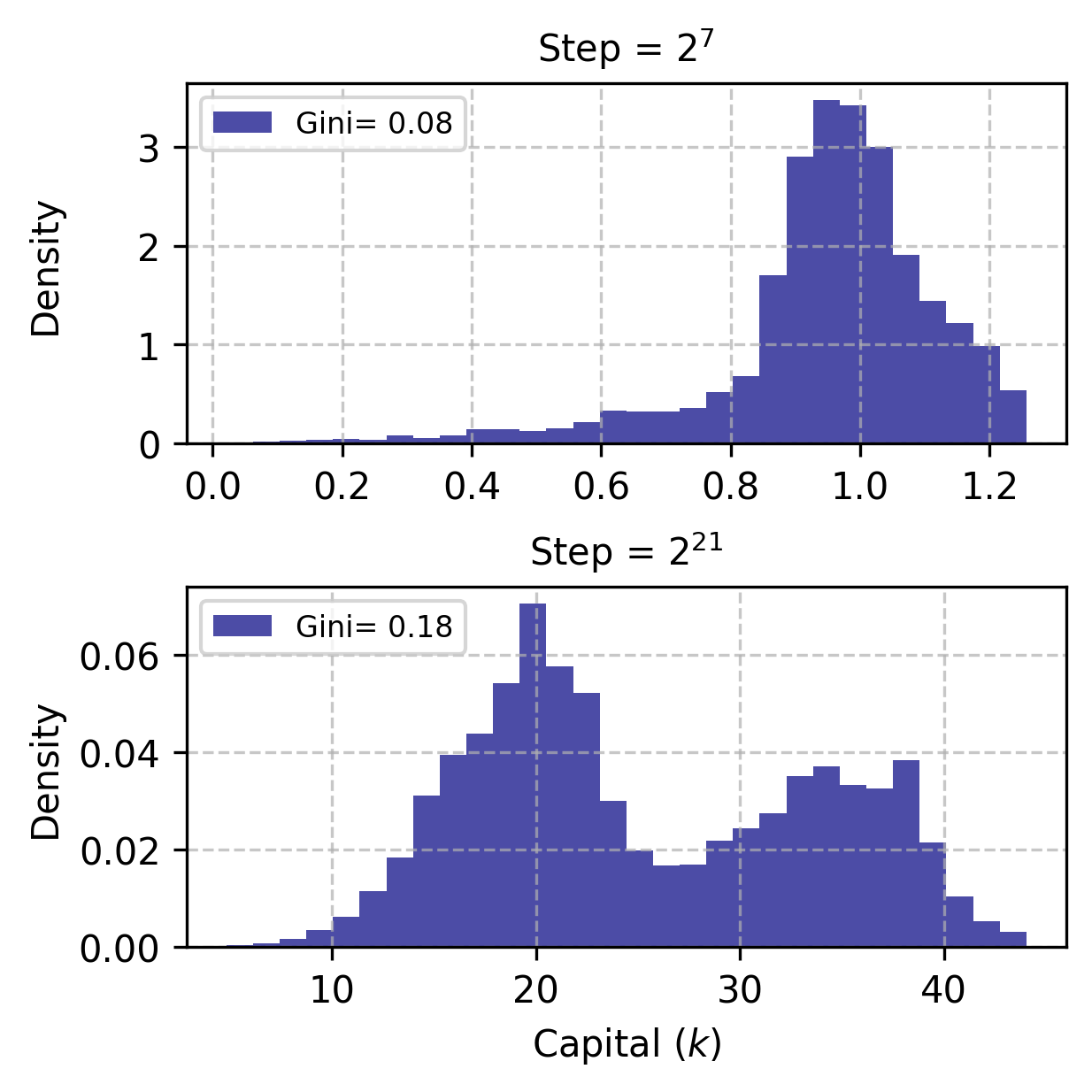}\hfill
  \includegraphics[width=0.33\textwidth]{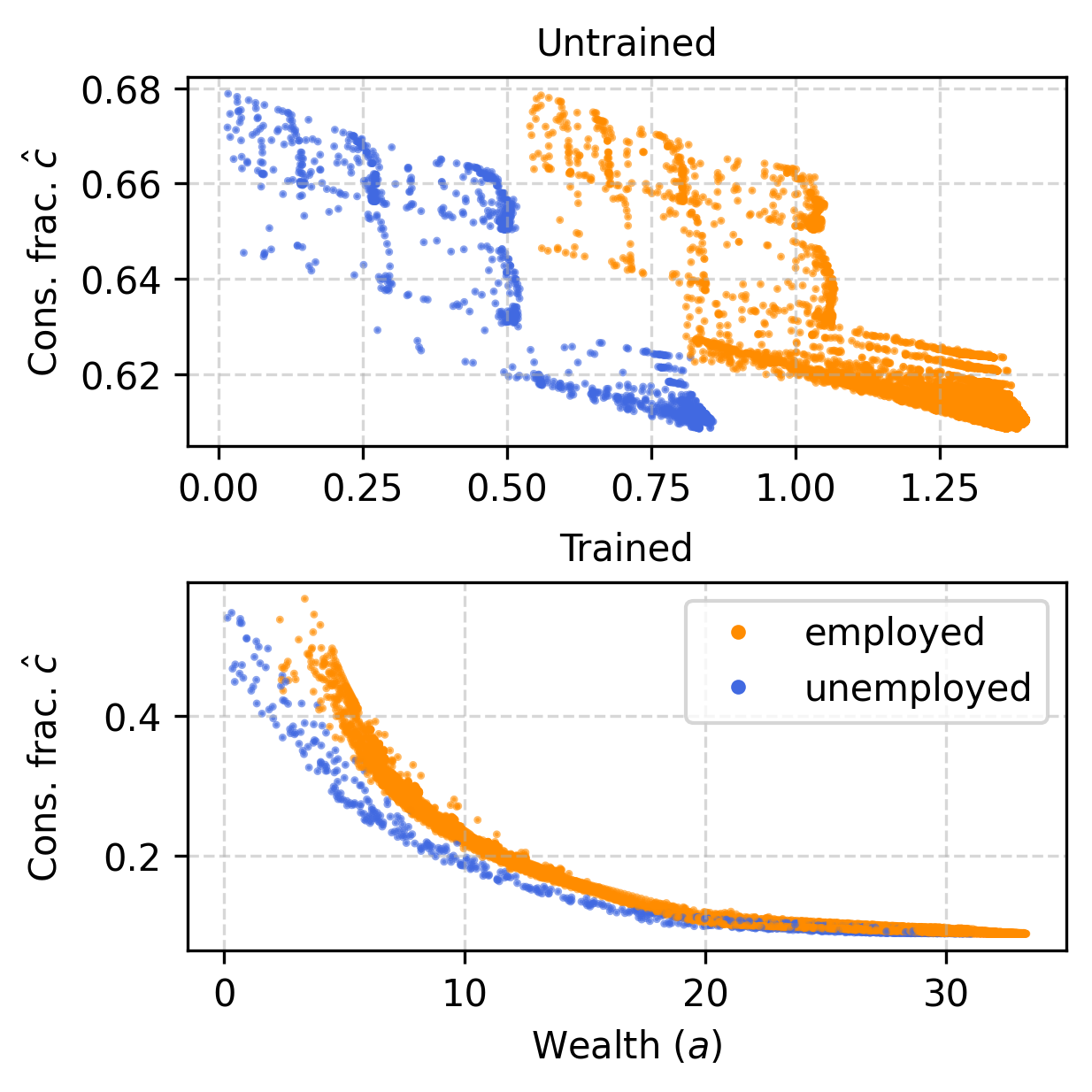}
  \caption{\algoName{} can reproduce mean‐field \textit{Krusell–Smith} results with a large number of identical agents.
  \normalfont
  The \textbf{left} panels are scatter plots of the aggregate capital at two consecutive times $K_{t}$ and $K_{t+1}$, at four training snapshots, with progressively more training steps (namely $2^7 \approx 10^2$, $2^{11} \approx 2\cdot10^{3}$, $2^{19} \approx 5\cdot10^{4}$ and $2^{21} \approx 2\cdot10^{6}$).
  The characteristic linear time evolution of aggregate capital (cf. the first figure of the original paper \cite{KrusellSmith1998}) is here recovered with a sufficient number of training steps.
  The \textbf{centre} panels depict the wealth distributions of the agents before and after training.
  While the Gini index for the untrained model is 0.08, it moves to 0.18 for the trained model distribution, a number which is in line with the original work.
  Finally, the \textbf{right} panels illustrate the curves of the marginal propensities to consume learned by the \algoName agents.
  Analogously to the KS agent policy, the consumption policy learned by the \algoName agents is almost completely flat when wealth is high, while it increases sharply when wealth is low.
  }
  \label{fig:KS_results}
\end{figure*}

\vspace{\secvspace}
\subsection{Representative agent RBC limit}
\label{subsec:results_rbc}

The classic RBC model entails a single representative household, assumed to internalise all aggregate variables.
As illustrated in the top right panel of Figure~\ref{fig:illustration}, our framework can recover exactly the RBC model as a special case by just choosing a single agent and unitary productivity of labour and capital, i.e.
\begin{equation} n=1,\qquad \kappa=\lambda=1, \end{equation}
so that individual and aggregate quantities coincide: $K_t=k_t$ and $L_t=\ell_t$.
Furthermore, as for all standard RBCs, we choose the technology to move according to the exponential of an AR(1) process
\begin{equation} A_t = e^{z_t}, \qquad z_{t} \;=\; \rho \,z_{t-1} + \sigma \varepsilon_{t}, \end{equation}
where $\rho$ and $\sigma$ are the persistence and volatility of the stochastic process, and $\varepsilon_t$ is a random number sampled independently from a standard normal distribution.

\vspace{\parvspace}
\noindent
\textbf{Algorithmic performance.}
Figure~\ref{fig:learning_curves} shows the four RL schemes considered (PPO, SAC, TD3 and DDPG) trained on the textbook RBC problem. 
All RL algorithms considered for this problem found successful policies to optimise cumulative rewards.
More specifically, SAC, TD3 and DDPG appear to strongly outperform PPO in terms of speed of convergence towards the optimal solution, with SAC being the most stable learner among all.
As we will show later (see Figure \ref{fig:scalability}), SAC is the strongest candidate for the multi-agent experiments.
DDPG is another valid option for the single-agent experiments since, although it takes slightly more steps to converge, it is also faster than SAC in terms of wall-clock time.
For these reasons, we choose to leverage the speed of DDPG in the RBC limit experiments shown here and the efficacy and stability of learning of SAC for the multi-agent experiments.

\vspace{\parvspace}
\noindent
\textbf{Reproducing textbook RBCs ($\boldsymbol{\delta=1}$).}
Here, we reproduce the results of the simplest type of RBC model as covered in most macroeconomic textbooks, one with full depreciation ($\delta=1$).
In this setting, capital cannot be transferred to the next step and closed-form solutions can be obtained \cite{brock1972optimal}.
The optimal policy $(\hat{c}^\star_t, \ell^\star_t)$ reads
\begin{equation} \hat{c}^\star_t \;=\; (1-\alpha\beta), \qquad \ell^\star_t \;=\; \frac{\alpha}{\,b\bigl(1-(1-\alpha)\beta\bigr)+\alpha}. \label{eq:textbook_rbc_actions} \end{equation}

The left panel of Figure~\ref{fig:rep_agent_results} shows how the RL household learns to recover the optimal actions of the textbook RBC.
Specifically, after around $10^4$ training steps, the consumption fraction $\hat{c}_t$ and labour $\ell_t$ (shown in blue and orange) approximately converge to the optimal values shown as dashed lines, and then remain stably around those values.

\vspace{\parvspace}
\noindent
\textbf{Reproducing typical RBCs ($\boldsymbol{\delta=0.025}$).}
More general RBCs set up a partial depreciation, often of $\delta=0.025$ as done in our experiments.
Relaxing the full-depreciation assumption makes the model impossible to solve analytically.  
In these cases, economic modellers typically resort to approximate solutions relying on first-order conditions and linearisation as implemented in widespread software such as `Dynare'~\cite{villemot2011solving}.
The centre panel of Figure~\ref{fig:rep_agent_results} shows that, after approximately $10^4-10^5$ training steps, the RL household of our framework learns optimal consumption and labour choices that are coherent with those computed with the Dynare software.

\vspace{\parvspace}
\noindent
\textbf{Reproducing impulse response functions.}
We further demonstrate that our framework can reproduce standard impulse response functions.
To compute them, we initialise the model at its deterministic steady state (computed with \(\sigma_z=0\)), hit the economy with a one-standard-deviation productivity shock, and observe the relaxation to the steady state value.
The right panel of Figure~\ref{fig:rep_agent_results} overlays the impulse–response functions for consumption from our framework (blue lines with the shaded area) and Dynare (black line); the two lines are statistically consistent, confirming that our framework correctly reproduces not only the stationary choices but also the dynamic response to external shocks.

\begin{figure*}[htbp]
  \centering
  \includegraphics[width=0.33\textwidth]{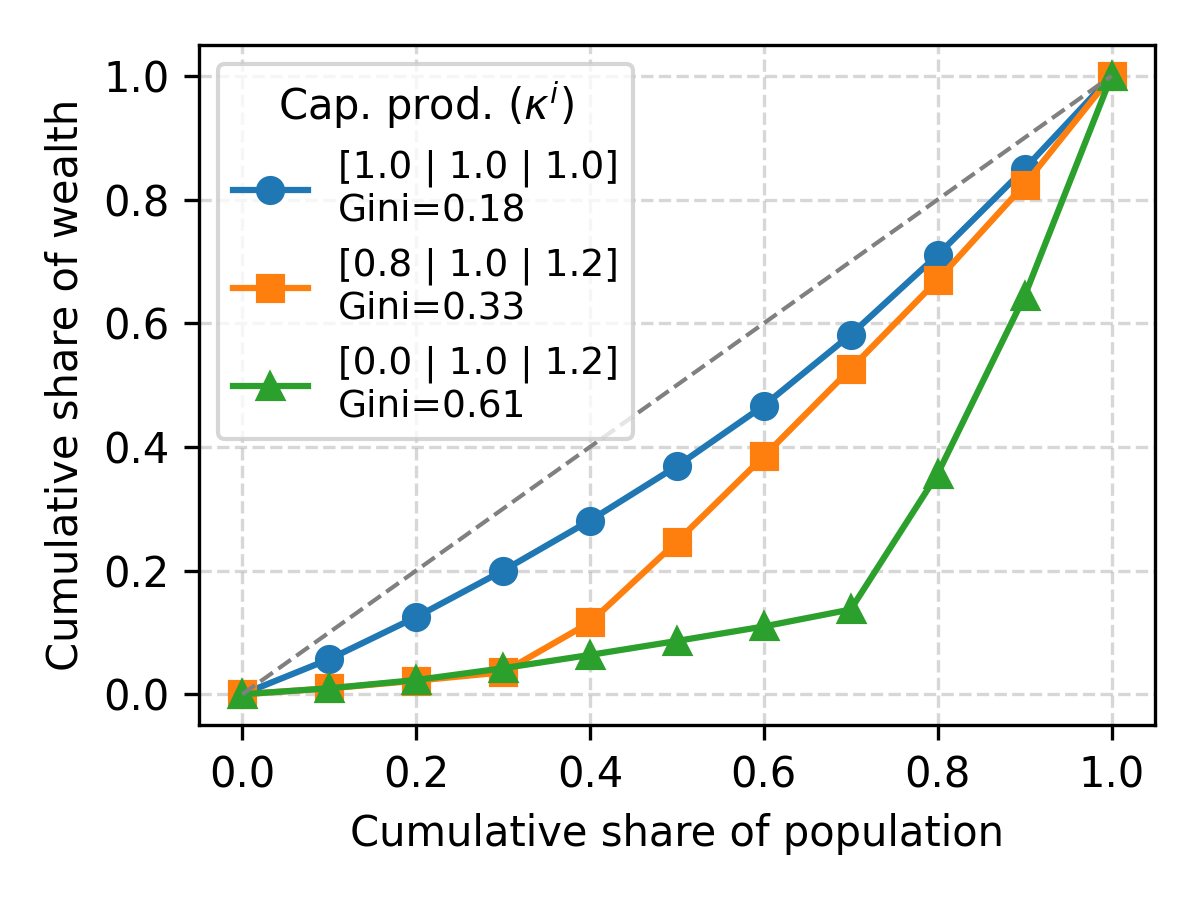}
  \includegraphics[width=0.33\textwidth]{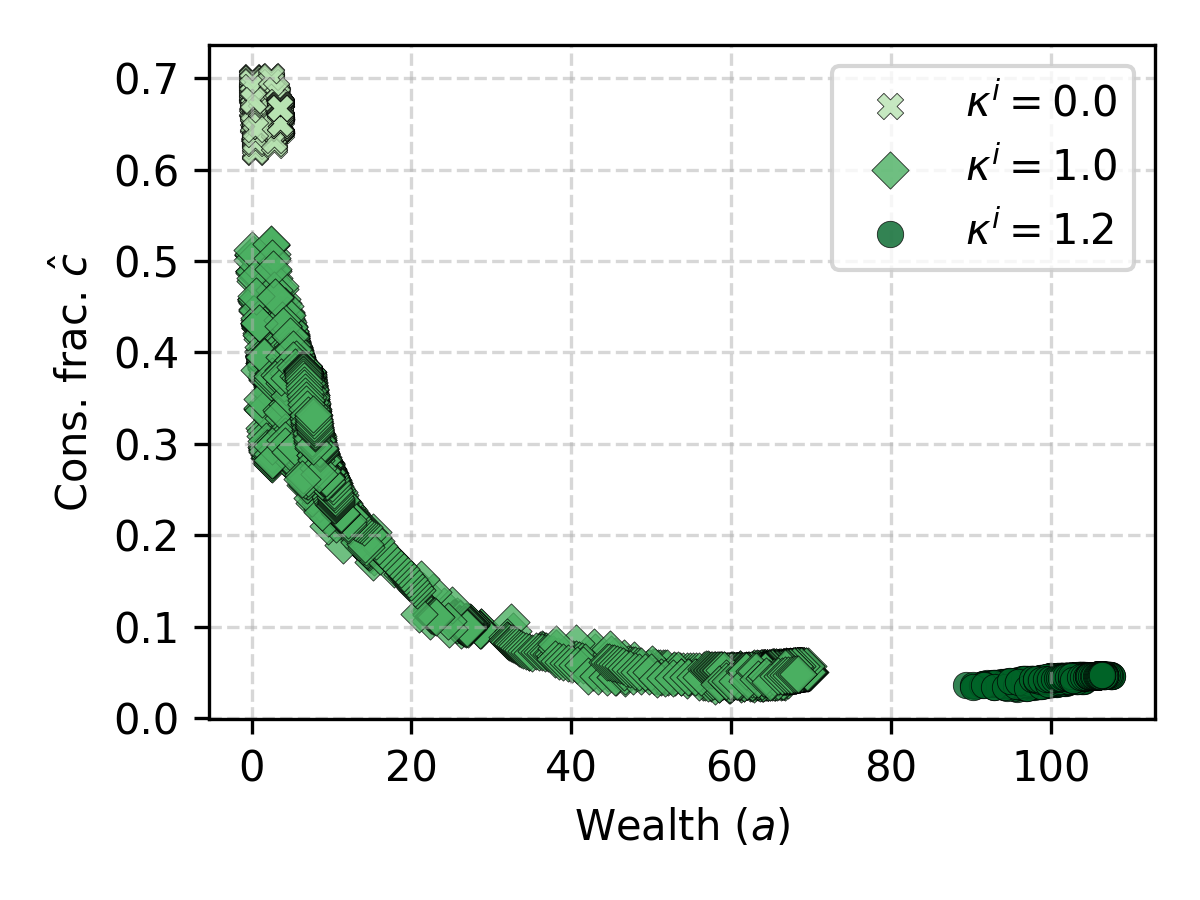}
  \includegraphics[width=0.33\textwidth]{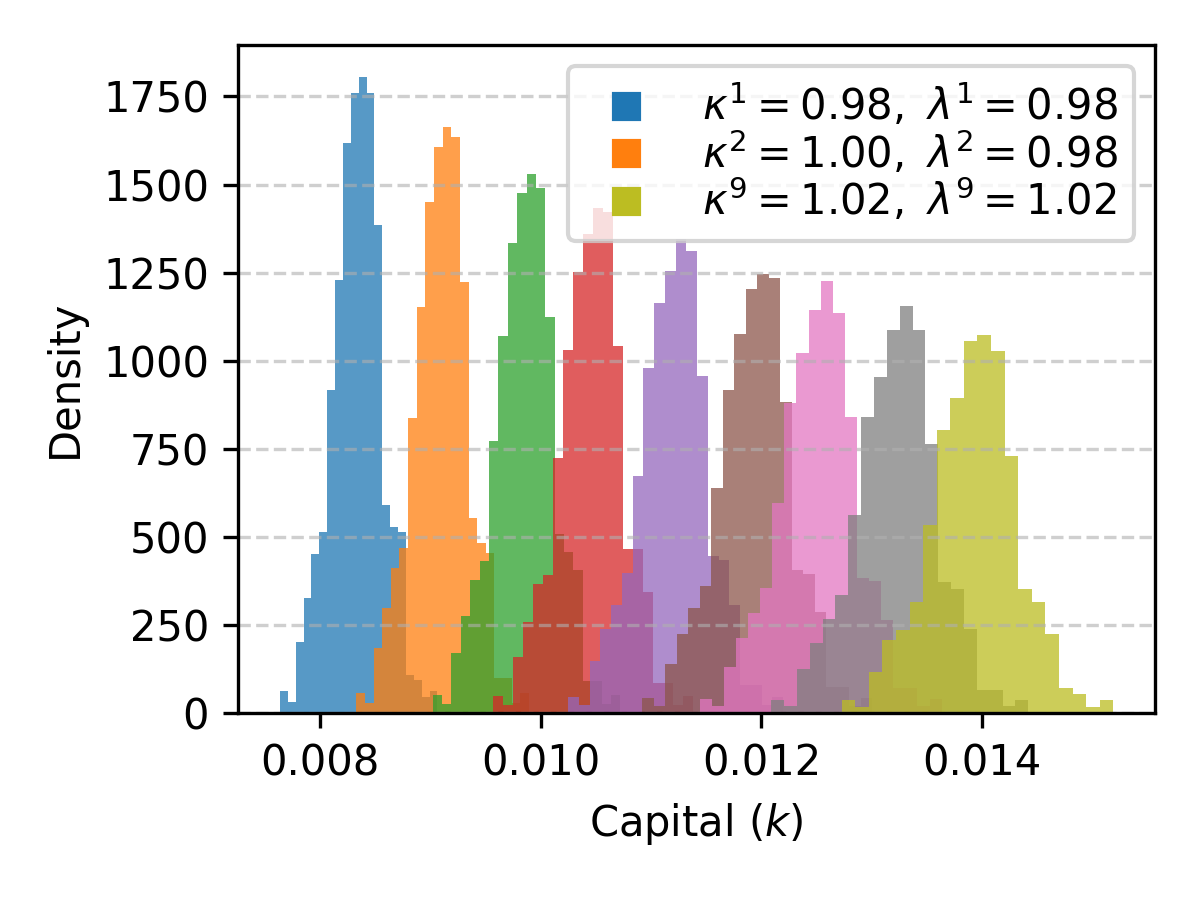}
  \caption{\algoName allows for modelling rich ex-ante heterogeneity and for the emergence of heterogeneous behaviour.
    \normalfont
    The \textbf{left} panel reports the Lorenz curves, and the corresponding Gini indices, of the wealth distributions computed using \algoName, either in the standard KS setting (blue dots) or in the extended KS with heterogeneous capital productivities (orange circles and green triangles).
    The graph illustrates how \algoName allows one to model a wider range of wealth inequalities.
    The \textbf{centre} panel reports the marginal propensities to consume of the three classes of agents (of low, medium and high returns) for the KS model with Gini index equal to 0.61.
    The graph illustrates how the three household groups learn markedly different policies.
    The \textbf{right} panel illustrates the kind of wealth distribution that one can obtain from an RBC model with 9 agents with different combinations of labour and capital productivity.
    The legend shows only 3 colours, but the pattern is clear: higher productivities imply higher wages or capital returns that, in turn, give rise to higher capital accumulation.
    }
  \label{fig:marlbc_results}
\end{figure*}

\vspace{\secvspace}
\subsection{Mean-field Krusell--Smith limit}
\label{subsec:results_ks}

The celebrated `Krusell--Smith' (KS) model \cite{KrusellSmith1998} is a mean-field model with ex-ante identical households that face discrete shocks of individual unemployment and aggregate technology. 
As illustrated in the bottom right panel of Figure~\ref{fig:illustration}, our framework can recover the mean-field KS limit with a large number of identical agents
\begin{equation} n \gg 1, \qquad \kappa^i=\lambda^i=1 \;\; \forall i. \end{equation}
In practice, in our experiments we use $n=20$ agents as we expect this to be sufficient to reproduce trends and distributions of the mean field limit.
Furthermore, to recover the precise KS model behaviour, we set up the aggregate and individual shocks in close agreement with the original model.
Specifically, we let the aggregate technology factor $A_t$ follow a two‐state Markov chain $A_t \in [A_t^{g}, A_t^{b}]$.
In the `good' ($g$) or `bad' ($b$) state, technology is respectively higher $A_t^{g} = 1.02$ or lower $A_t^{b} = 0.98$.
Furthermore, individual households also follow a two-state Markov chain, being either employed and supplying $\ell^i_t = \bar\ell = 1.11$ units of labour, or being unemployed and supplying no labour ($\ell^i_t = 0$).
The transition probabilities between aggregate and individual states are taken from \cite{KrusellSmith1998}.
They are calibrated to US data to make the expected duration of each regime eight periods, and in such a way that the unemployment rate is around $4\%$ in good states and $10\%$ in bad states. 
Since labour is fixed, we let the RL household choose only the consumption fraction $\hat{c}^i_t$ and set $b=0$ in the reward function.
Finally, in agreement with \cite{KrusellSmith1998}, we set the observation space to $\mathbf{x}^i_t=(k^i_t,\ell^i_{t},K_t,A_t)$.
The results of these experiments are given in Figure~\ref{fig:KS_results} through a single illustrative training run, and discussed below.
Statistics over different seeds are available in \cite[Appendix]{gabriele2025heterogeneous_ARXIV}.

\vspace{\parvspace}
\noindent
\textbf{Reproducing the classic KS law of motion.}
The left panel of Figure~\ref{fig:KS_results} illustrates the convergence properties of our \algoName framework relative to the classic `law of motion' for aggregate capital postulated in the original paper \cite{KrusellSmith1998}.  
The four small scatter plots display the pairs
$(K_t,K_{t+1})$ after approximately $10^2$, $2\cdot10^{3}$, $5\cdot10^{4}$, and $2\cdot10^{6}$ SAC updates.
In the KS procedure, a linear map is postulated whose coefficients are estimated until they converge in a self-consistent procedure.
By contrast, using \algoName{} a perfectly linear relationship ($R^{2}>0.99$) between consecutive aggregate capitals emerges endogenously without any a priori assumption.

\vspace{\parvspace}
\noindent
\textbf{Reproducing KS distributional properties.}
The two histograms in the centre panel of Figure~\ref{fig:KS_results} report the stationary distribution of individual capital before and after convergence.
Before convergence, the scale of the capital distribution is very small, ranging only from 0 to 1.2. This is a sign that the agents have not yet learned consumption policies capable of accumulating capital. Not surprisingly, the Gini index is very low.
After convergence, the Gini index increases to $0.18$, a value relatively close to the 0.25 value from the original KS calculation.

\vspace{\parvspace}
\noindent
\textbf{Reproducing KS marginal propensity to consume.}
The two graphs in the right panel of  Figure~\ref{fig:KS_results} report the agents' marginal propensity to consume as a function of their wealth, before and after training (top and bottom graphs).
The graphs show that the learning process converges to curves that are flat for large values of wealth ($a>20$) and increase rapidly for low values of wealth (especially after $a<5$).
Interestingly, there is very little difference in the consumption fractions if the agents are employed (orange dots) or unemployed (blue dots).
These shapes for the curves of the marginal propensity to consume are another key result of the original KS paper that the \algoName can easily recover.

\vspace{\secvspace}
\subsection{Greater heterogeneity with \algoName}
\label{subsec:results_general}

We illustrate here how \algoName can seamlessly extend classic models by leveraging its flexibility to represent a greater level of agent heterogeneity.
Specifically, we extend the KS model with heterogeneous capital productivities and then extend the RBC model with heterogeneity in both capital and labour productivities.

\vspace{\parvspace}
\noindent
\textbf{KS with heterogeneous capital returns.}
We extend the KS model by dividing the $n=20$ agents into three groups, respectively with low, middle and high capital productivity.
Specifically, we assign 14 agents (or 70\% of the total) to the middle productivity group, and 6 agents (or 30\% of the total) equally between the low and high productivity groups.
Although this assignment is largely arbitrary, the described proportions were selected to resemble those of a normal distribution, where roughly 70\% of the data falls within one standard deviation of the mean and 30\% lies in the two tails.
We perform two experiments, with either mild or marked differences in capital productivity among the three groups.
To be more precise, in the `mild' experiment we let the three productivities $\kappa^i$ be $\{ 0.8, 1.0, 1.2 \}$, while in the `marked' experiment we let them be $\{ 0.0, 1.0, 1.2\}$.
As capital productivities have an immediate effect on capital returns via Eq.~\eqref{eq:returns_and_wages}, this KS extension can neatly represent the heterogeneous returns on capital of real economies \cite{xavier2021wealth}.

The left panel of Figure~\ref{fig:marlbc_results} reports the Lorenz curves, and the corresponding Gini indices, of the original KS experiment (blue dots) as well as of the mildly heterogeneous experiment (orange squares) and markedly heterogeneous experiment (green triangles). 
The graph nicely illustrates how the introduction of heterogeneous capital productivity, and hence of heterogeneous capital returns, allows for the modelling of more unequal economies, with Gini indices increasing to 0.33 and 0.61 in the mildly and markedly heterogeneous returns models.
Different capital returns also give rise to different marginal propensities to consume as reported in the centre panel of Figure~\ref{fig:marlbc_results}.
Specifically, the graph shows a scatter plot of the consumption fraction as a function of wealth for the markedly heterogeneous experiment.

The graph shows that the households with lower capital returns (green crosses) learn a policy known as `hand-to-mouth', in which they consume almost all of their wealth each step without any capital accumulation.
This type of policy is often encoded in GE models as a heuristic, while it emerges endogenously with our framework as a result of learning.
On the other extreme, households with higher capital returns learn to consume a very low portion of their wealth since this policy leads to wealth accumulation and thus to higher future consumption.

\vspace{\parvspace}
\noindent
\textbf{RBC with heterogeneous returns and wages.}
To further illustrate the flexibility and scope of \algoName, we build a heterogeneous version of the standard RBC described in Sec.~\ref{subsec:results_rbc} with nine agents, each with capital and labour productivity taken from a $3\times3$ grid containing values in the set $\{ 0.98, 1, 1.02\}$.
Thus, the agent with the lowest productivity has $\kappa^i=\lambda^i=0.98$, the one with the highest productivity has $\kappa^i=\lambda^i=1.02$, and the other agents have all combinations in the middle.
In turn, these different productivities give rise to heterogeneity in both capital returns and wages, which influence consumption and labour decisions as well as wealth levels. 
In the right panel of Figure~\ref{fig:marlbc_results}, we illustrate the results of this setup by showing how the nine agents stabilise with overlapping yet distinct levels of wealth after a sufficient number of training steps.
Although we do not show this in the figure, the different productivities also give rise to significantly different choices for consumption fraction and labour supply.

\begin{figure}[t]
  \includegraphics[width=0.99 \linewidth, trim=0.2cm 0cm 0cm 0cm, clip]{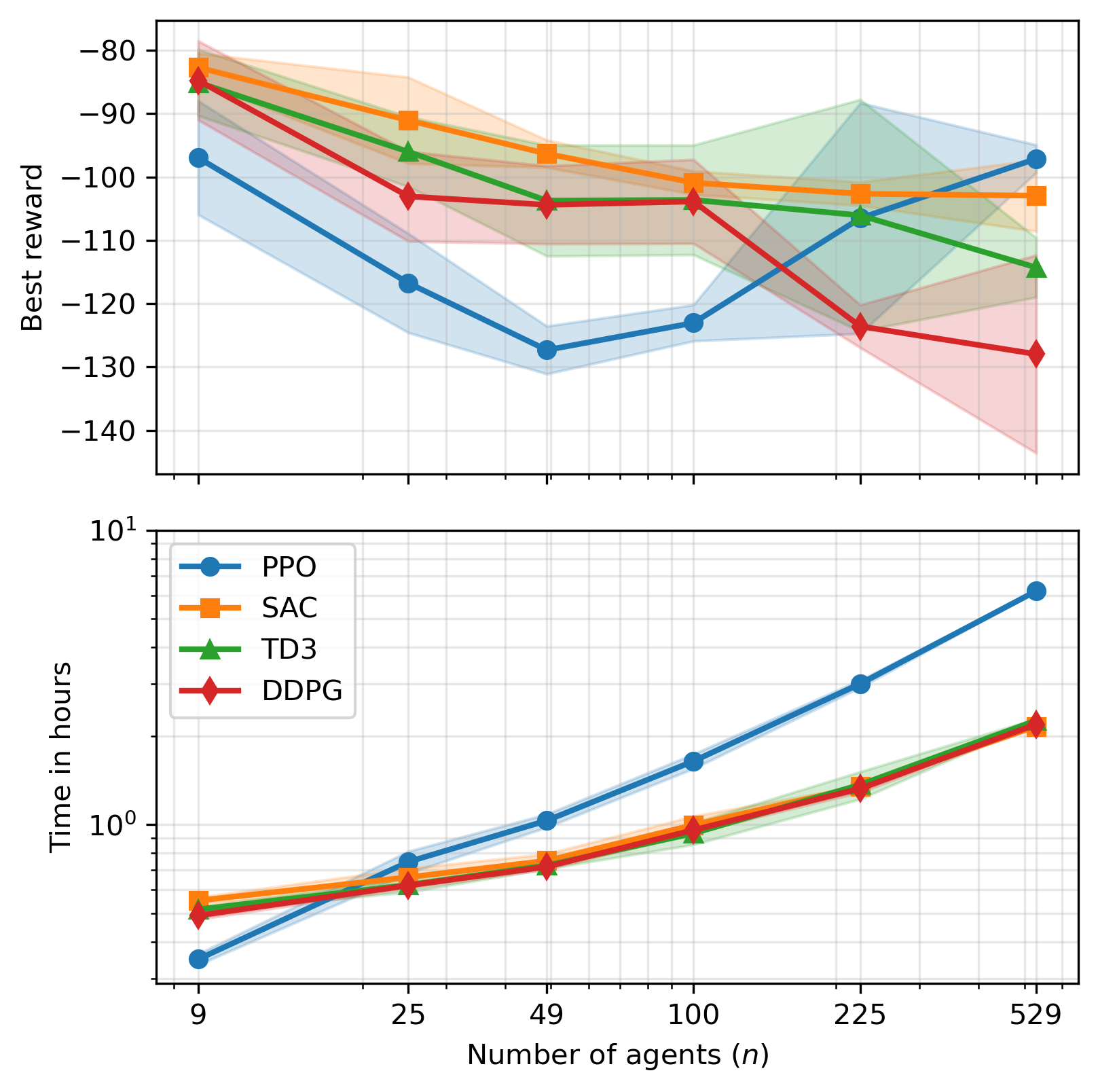}
  \caption{
  \algoName can scale up to hundreds of agents.
  \normalfont
  The \textbf{top} panel reports the best rewards achieved by \algoName agents during training runs of $10^{5}$ per-agent updates, meaning $n \cdot 10^{5}$ steps in total. 
  The \textbf{bottom} panel reports the time needed for the corresponding runs, on a single CPU machine.
  Both quantities are shown as a function of the number of agents ($n$) in the model.
  PPO is seen to underperform with respect to its competitors. 
  On the contrary, SAC is observed to obtain comparatively high rewards, in reasonable time limits, even for the largest models considered.
  }
  \label{fig:scalability}
\end{figure}

\vspace{\parvspace}
\noindent
\textbf{Scalability to hundreds of agents.}
To test the scalability of \algoName to simulations with a large number of heterogeneous agents, we expand on the experiment discussed in the previous paragraph and perform training runs for progressively denser, equispaced, grids of capital and labour productivities ($\kappa^i$ and $\lambda^i$), from $3\times3=9$ agents to  $23\times23=529$ agents.
The results of these runs are shown in Figure~\ref{fig:scalability}.
The top panel in the figure reports the mean and two standard deviations of the best episodic return attained during training runs of $10^{5}$ per-agent updates.
The bottom panel reports the corresponding wall-clock time taken for training on a single CPU. 
We find that the performance of SAC remains stably high across all model sizes.
On the contrary, we find that PPO and DDPG strongly underperform respectively in regimes of low and high numbers of agents.
Finally, we find that the computational cost rises steadily with the number of agents, and yet remains within practical limits for SAC (it takes roughly two hours to train the largest model for roughly $50\cdot10^6$ steps).
These results indicate that \algoName can be successfully trained with hundreds of households, even using very modest hardware.
Further details on the scalability of the framework are available in \cite[Appendix]{gabriele2025heterogeneous_ARXIV}.

\vspace{\Secvspace}
\section{Conclusions}
\label{sec:conclusions}

In this work, we present the Multi-Agent Reinforcement Learning Business Cycle (\algoName) framework, a macroeconomic modelling framework that integrates deep MARL within a real business cycle (RBC) environment.
We show that \algoName can reproduce the canonical representative-agent RBCs and the Krusell–Smith mean-field models, and it can extend them by simulating multiple heterogeneous agents. 

One obvious limitation of our framework is the computational cost, with accurate training of RL agents requiring up to a few hours to terminate successfully.
However, the use of vectorised GPU computation and of more modern MARL schemes is likely to greatly mitigate this shortcoming in the near future.
Specifically, by implementing the MARL environment described in a GPU-compatible form, for instance using specialised software~\cite{flair2024jaxmarl}, would allow one to fully leverage GPU acceleration (which can easily exceed a factor of 20).
Furthermore, while the independent learning scheme used here has the advantage of being simple to understand and implement, more sophisticated MARL approaches could improve learning stability and speed \cite{albrecht2024multi}.
This would allow much faster and more robust training, hence greatly facilitating usability, prototyping and extension of \algoName.

While this study focused primarily on the description of \algoName, its limit cases, and its capabilities, future work could investigate the use of the framework to study specific economic problems.
Natural issues could be related to economic inequality, but also to the economic and financial consequences of asymmetric changes in labour productivity, such as those potentially deriving from the spreading of AI tools in the workplace.
By showing how a model of multiple interacting agents (akin to an ABM) can give rise to heterogeneous GE results, our work marks a step towards a synthesis of these often opposed methodologies.
Furthermore, by showing how a MARL approach can recover and extend classic results from the economics literature, our work has significant potential to foster communication and exchanges between communities in computer science and economics that are interested in RL modelling, but that have not previously found common ground.

\begin{acks}
We would like to thank Alessandro Moro, Valerio Astuti, Valerio Nispi Landi, Sergio Santoro (Banca d'Italia), Doyne Farmer and José Morán (Oxford University), Marta Grześkiewicz (Cambridge University) and Sara Casella (LUISS) for useful feedback, and Michele Colombi (Scuola Normale Superiore) for double-checking some of the results.
\end{acks}

\bibliographystyle{ACM-Reference-Format}
\bibliography{sample}

\balance


\clearpage
\newpage
\onecolumn

\setcounter{section}{0}
\setcounter{subsection}{0}
\setcounter{subsubsection}{0}

\appendix

\vspace{\Secvspace}
\section{Appendix}

We provide here additional details and experimental results complementing the information reported in the main text. 
First, in Sec.~\ref{SM:hyperparams}, we document the hyperparameter settings used for each reinforcement learning (RL) algorithm to improve reproducibility.
Then, in Sec. \ref{SM:extra_results}, we report extended quantitative results on the learning process for the multi-agent experiments already described in the main text to further illustrate learning convergence and, in turn, the robustness and scalability of \algoName to large population sizes.

\vspace{\secvspace}
\subsection{Hyperparameters of the RL algorithms}
\label{SM:hyperparams}
In the interest of reproducibility, we provide here further details on the hyperparameters of the RL algorithms used.
All experiments were implemented using the default architectures provided in the Stable Baselines 3 library (v2.5.0). 
%
PPO employs a shared multilayer perceptron for policy and value estimation, with two 64-unit \texttt{tanh} layers, following the standard on-policy actor–critic architecture \cite{schulman2017proximal}.
SAC uses separate actor and critic networks, each with two 256-unit \texttt{ReLU} layers, optimising a stochastic policy under an entropy-regularised objective \cite{haarnoja2018softactorcriticoffpolicymaximum}.
TD3 and DDPG adopt two-hidden-layer networks (400 and 300 units, \texttt{ReLU}) for both actor and critics; TD3 additionally trains two critics independently and delays policy updates to mitigate value overestimation \cite{fujimoto2018td3, lillicrap2016ddpg}.
Table~\ref{tab:rl_hparams} summarizes the main learning hyperparameters adopted across algorithms in the experiments.
%

\vspace{\parvspace}
\noindent
\textbf{Code availability.}
We repeat here that the code used to generate the results is available at \url{https://github.com/fedegabriele/MARL-BC}.

\begin{table*}[th]
\centering
\begin{minipage}[t]{0.24\textwidth}
\centering
\caption*{PPO}
\begin{tabular}{ll}
\hline
\textbf{Parameter} & \textbf{Value} \\
\hline
learning rate & 3e-4 \\
n steps & 2048 \\
batch size & 64 \\
n epochs & 10 \\
gae lambda & 0.95 \\
clip range & 0.2 \\
ent coef & 0.0 \\
vf coef & 0.5 \\
max grad norm & 0.5 \\
\hline
\end{tabular}
\end{minipage}
\hfill
\begin{minipage}[t]{0.24\textwidth}
\centering
\caption*{SAC}
\begin{tabular}{ll}
\hline
\textbf{Parameter} & \textbf{Value} \\
\hline
learning rate & 3e-4 \\
buffer size & 1e6 \\
learning starts & 100 \\
batch size & 256 \\
tau & 0.005 \\
ent coef & learned \\
target entropy & learned \\
\hline
\end{tabular}
\end{minipage}
\hfill
\begin{minipage}[t]{0.24\textwidth}
\centering
\caption*{TD3}
\begin{tabular}{ll}
\hline
\textbf{Parameter} & \textbf{Value} \\
\hline
learning rate & 1e-3 \\
buffer size & 1e6 \\
learning starts & 100 \\
batch size & 256 \\
tau & 0.005 \\
policy delay & 2 \\
target policy noise & 0.2 \\
target noise clip & 0.5 \\
\hline
\end{tabular}
\end{minipage}
\hfill
\begin{minipage}[t]{0.24\textwidth}
\centering
\caption*{DDPG}
\begin{tabular}{ll}
\hline
\textbf{Parameter} & \textbf{Value} \\
\hline
learning rate & 1e-3 \\
buffer size & 1e6 \\
learning starts & 100 \\
batch size & 256 \\
tau & 0.005 \\
\hline
\end{tabular}
\end{minipage}
\vspace{10pt}
\caption{Hyperparameters for PPO, SAC, TD3, and DDPG.
\normalfont
Main hyperparameters for the four reinforcement learning algorithms used—Proximal Policy Optimization (PPO), Soft Actor-Critic (SAC), Twin Delayed Deep Deterministic Policy Gradient (TD3), and Deep Deterministic Policy Gradient (DDPG)—as implemented in Stable-Baselines3. 
%
}
\label{tab:rl_hparams}
\end{table*}

\newpage

\vspace{\secvspace}
\subsection{Extra results}
\label{SM:extra_results}

We report here additional experimental results complementing those described in Sections 4.2 and 4.3 of the main text.

\vspace{\parvspace}
\noindent
\textbf{Extra results related to section `4.2 Mean-field Krusell--Smith limit'.}

\begin{figure*}[h!]
    \centering
    \includegraphics[width=0.9\linewidth]{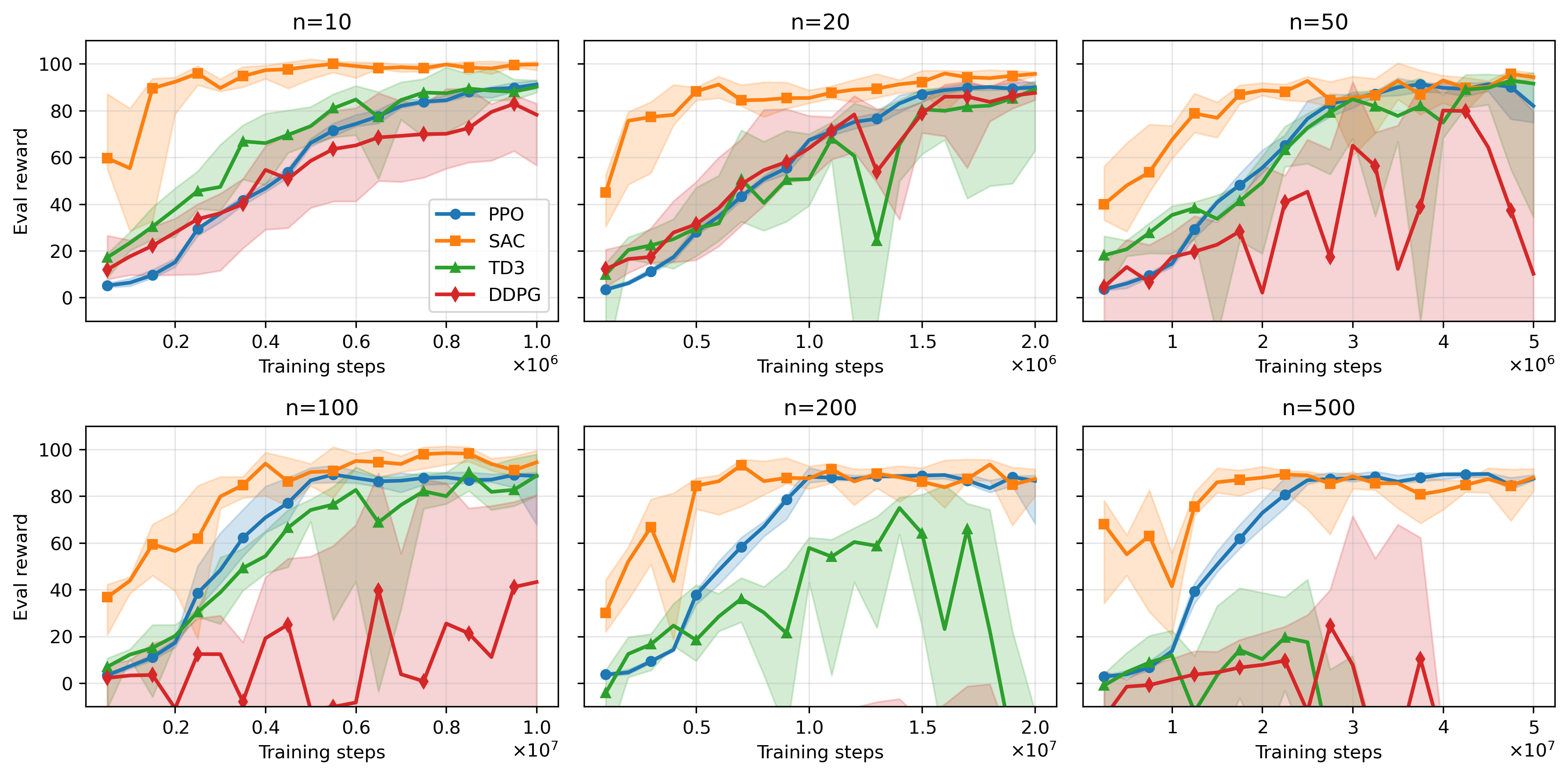}
    \caption{Learning curves for mean-field Krusell--Smith limit experiment. 
    \normalfont
    Each panel shows the evaluation reward as a function of training steps for a specific number of agents ($n$ = 10, 20, 50, 100, 200, 500). 
    Agents are ex-ante identical as in the mean-field (Krusell--Smith) model.
    The four RL algorithms are compared. 
    Solid lines represent the median reward across 8 independent runs, and the shaded region represents the range between 25th and 75th percentiles. 
    All panels share the same y-axis scale to facilitate direct comparison.
    }
\label{fig:learning_curves_KS}
\end{figure*}

\begin{figure}[h!]
    \centering
    \includegraphics[width=0.7\linewidth]{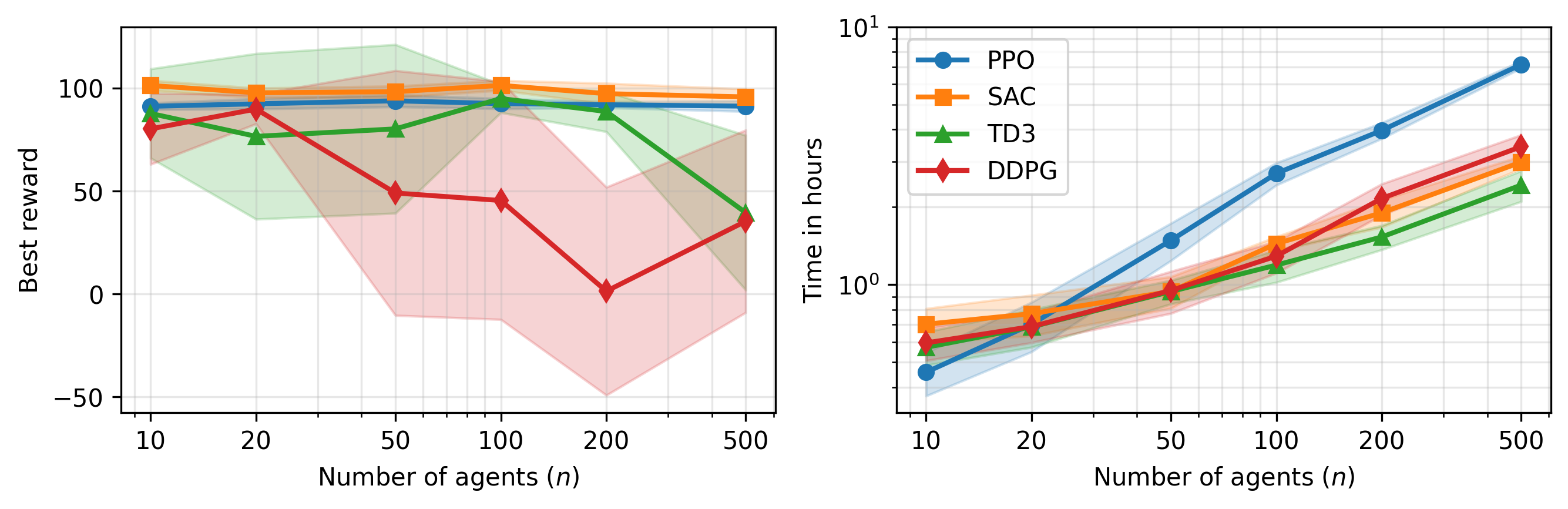}
    \caption{\algoName scaling to hundreds of agents for the mean-field Krusell--Smith limit. 
    \normalfont
      The \textbf{left} panel reports the best evaluation rewards achieved by \algoName agents during training runs of $10^{5}$ per-agent updates, meaning $n \cdot 10^{5}$ steps in total. 
      The \textbf{right} panel reports the time needed for the corresponding runs, on a single CPU machine.
      Both quantities are shown as a function of the number of agents ($n$) in the model.
      Solid lines and shaded areas report mean $\pm$ one standard deviation across the 8 training runs.
      SAC is observed to obtain comparatively high rewards, in reasonable time limits, even for the largest models considered.
    }
    \label{fig:scaling_KS}
\end{figure}

Figure \ref{fig:learning_curves_KS} shows the learning curves of the four RL algorithms for increasing agent numbers (from $n=10$ to $n=500$) in the mean-field Krusell--Smith limit of \algoName.
The figure shows that SAC consistently achieves high evaluation rewards across all population sizes.
Specifically, for relatively small agent populations $ n \le 20 $, SAC clearly outperforms any competing scheme while, as $n$ increases, PPO starts achieving similar rewards, although at a significantly slower pace.
TD3 and DDPG underperform for all population sizes, particularly in the large-$n$ regime where they fail to learn stably.
Importantly, the asymptotic reward level of SAC remains largely independent of $n$, indicating that the aggregate behaviour of the learned policy converges to the same mean-field equilibrium regardless of the population size.
This result is corroborated by the left panel of Figure \ref{fig:scaling_KS}, which directly shows the maximum reward achieved as a function of population size.
The right panel of the same figure confirms that SAC can be trained effectively within a reasonable time frame even on very modest hardware, as already demonstrated in Figure \ref{fig:scalability} of the main text.

\begin{figure}[h!]
    \centering
    \includegraphics[width=0.99\linewidth]{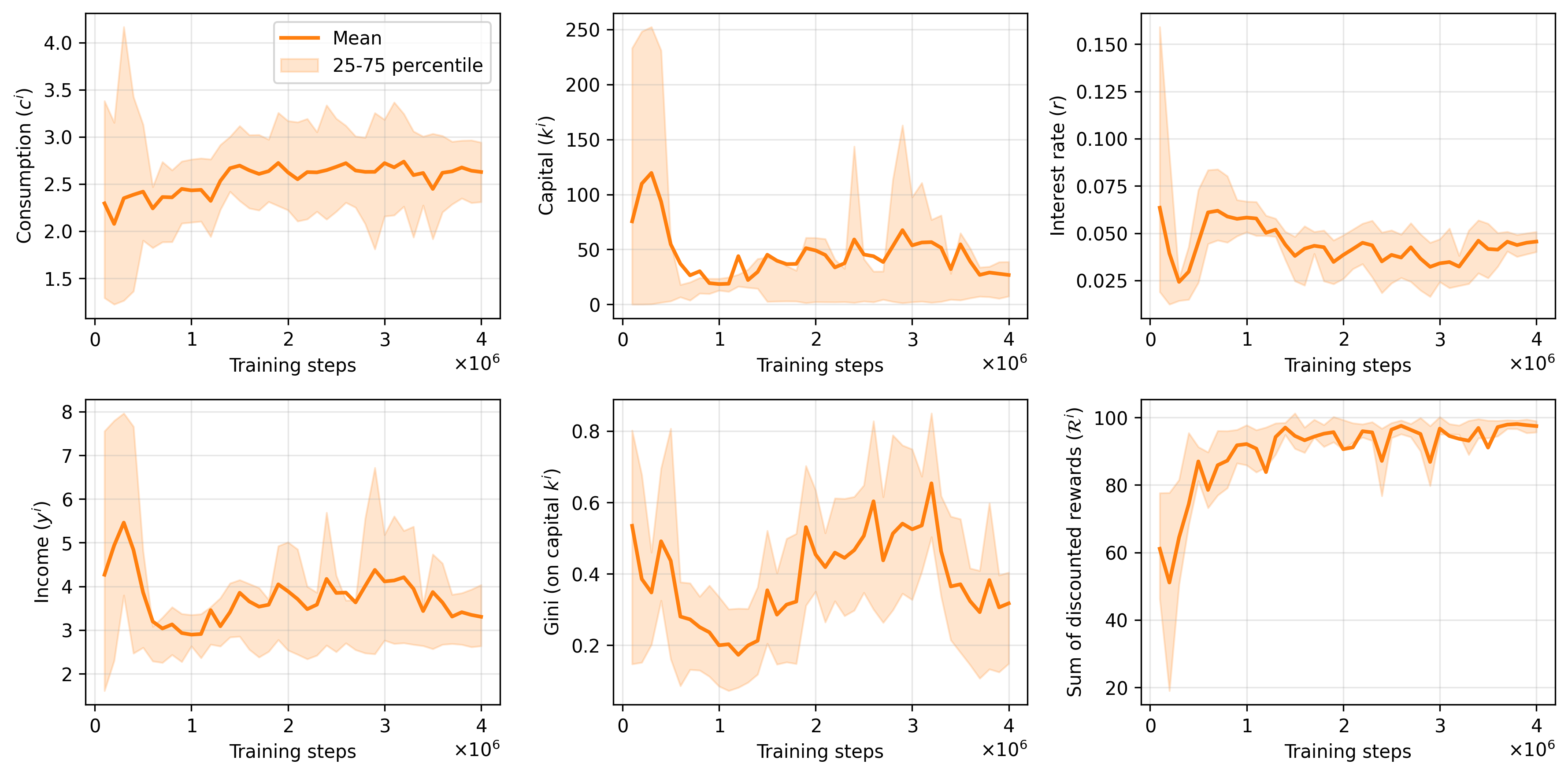}
    \caption{Training dynamics of the mean-field Krusell--Smith limit experiment with $n=20$ SAC agents.
    \normalfont
    From left to right, top to bottom, the six panels show: individual consumption ($c^i$), individual capital ($k^i$), interest rate ($r$), individual income ($y^i$), Gini index over capital, and the individual sum of discounted returns ($\mathcal{R}^i$).
    The individual quantities are computed by first averaging over the agents, and the lines show the mean and percentiles over eight training runs.
    }
    \label{fig:gini_stats_KS}
\end{figure}

Figure \ref{fig:gini_stats_KS} shows the evolution of a number of quantities during training in the mean-field Krusell--Smith limit experiment.
We note that as training progresses, the mean sum of discounted rewards ($\mathcal{R}^i$, bottom right panel) progressively grows and stabilises only slightly below $\mathcal{R}^i\approx 100$. 
In turn, this induces a state with consumption $c^i \approx 2.7$, capital $k^i \approx 25$, interest rate $r \approx 0.03$, and income $y^i \approx 3.3$, and with a Gini index on capital in the range $0.17-0.4$.

      %

\clearpage
\newpage
\vspace{\parvspace}
\noindent
\textbf{Extra results related to section `4.3 Greater heterogeneity with MARL-BC'.}

\begin{figure*}[h!]
    \centering
    \includegraphics[width=0.9\linewidth]{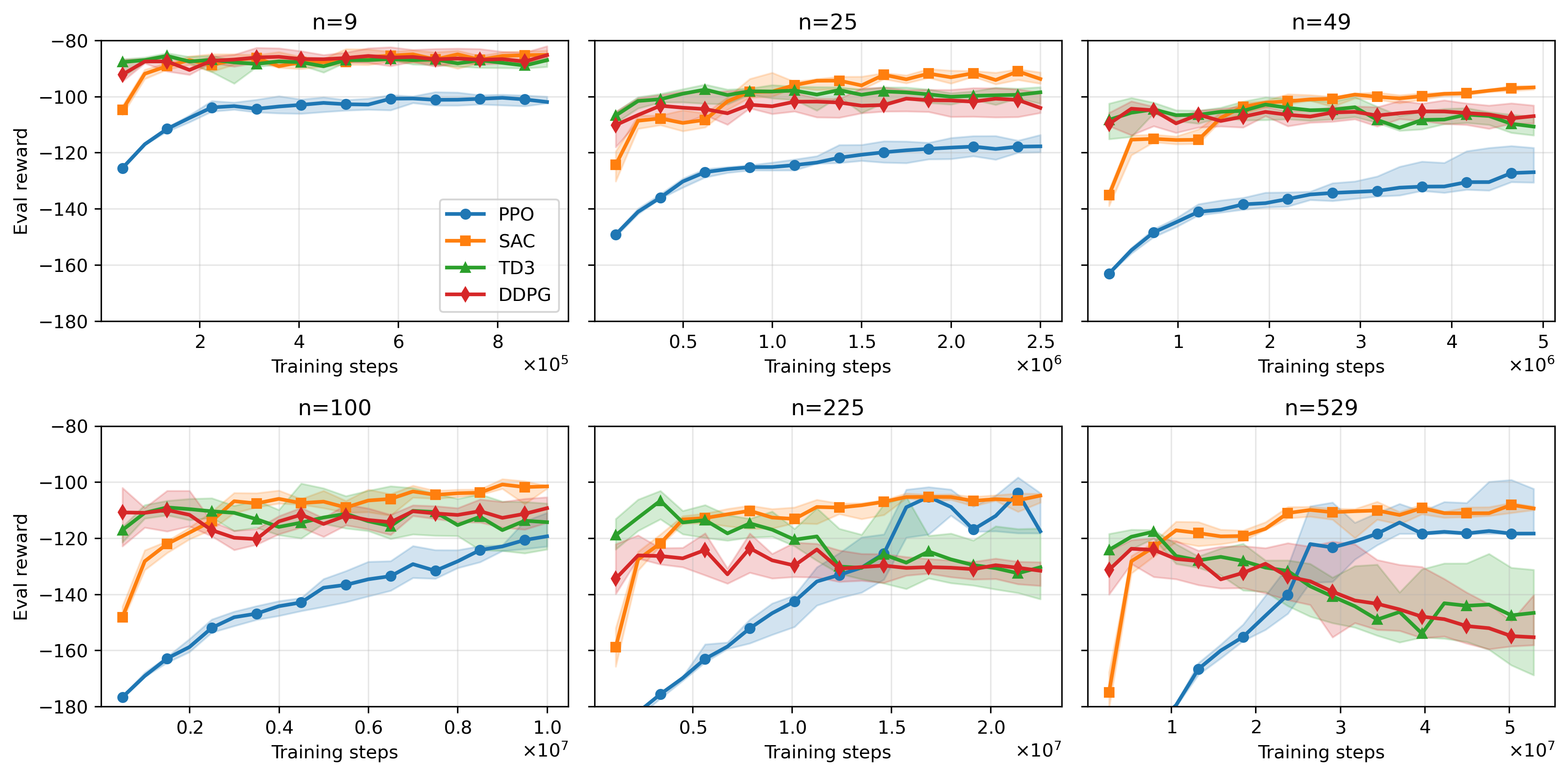}
    \caption{Learning curves for experiment on the RBC with grid-heterogeneous agents. 
    \normalfont
    Each panel shows the evaluation reward as a function of training steps for a specific number of agents ($n$ = 9, 25,49, 100, 225, 529). 
    Agents have ex-ante different capital and labour productivities taken from an equispaced $\sqrt{n} \times \sqrt{n}$ grid ranging from 0.98 to 1.02.
    The four RL algorithms are compared.
    Solid lines represent the median reward across 8 independent runs, and the shaded region represents the range between 25th and 75th percentiles.
    All panels share the same y-axis scale to facilitate direct comparison.
    }
    \label{fig:learning_curves_KL_HET}
\end{figure*}

Figure \ref{fig:learning_curves_KL_HET} shows the learning curves of the four RL algorithms for increasing agent numbers (from $n=9$ to $n=529$) in the RBC model with heterogeneous agents with capital and labour productivities taken from an equispaced two-dimensional grid.
For small populations $n \le 20$, SAC, TD3, and DDPG rapidly achieve high evaluation rewards, while PPO converges more slowly and to lower values.
As the number of agents grows, SAC remains the a very reliable learner, TD3 and DDPG show increasing instability and variance for large $n$.
PPO, despite slower learning, improves performance for increasing $n$ exhibiting stable convergence for the largest population sizes considered.
Importantly, the best attainable reward gradually converges toward a limiting value, suggesting that the heterogeneous system approaches a well-defined aggregate equilibrium for sufficiently large agent populations.
This result is already illustrated in the top panel of Figure \ref{fig:scalability} of the main text, which directly reports the best rewards achieved as a function of population size for this model.

\end{document}